\documentclass{elsart}

 \usepackage{graphics}

\usepackage{graphicx}

\usepackage{epsfig}

\usepackage{amssymb}

\begin{document}

\def\msun{${\rm M_{\odot}} \;$}
\def\be{\begin{equation}}
\def\ee{\end{equation}}
\def\gcc{g cm$^{-3}$}
\def\cc{cm$^{-3}$}
\def\bi{\begin{itemize}}
\def\ei{\end{itemize}}
\newcommand{\ii}{\item}
\def\ben{\begin{enumerate}}
\def\een{\end{enumerate}}

\def\i{\item}
\def\bea{\begin{eqnarray}}
\def\eea{\end{eqnarray}}
\def\p{\partial}
\def\dt{\Delta t}
\def\divv{\nabla \cdot \vec{v}}

\begin{frontmatter}















\title{Conservative, special-relativistic smoothed particle hydrodynamics}





\author{Stephan Rosswog}
\address{School of Engineering and Science, Jacobs University Bremen, 28759
  Bremen, Germany}

\begin{abstract}
We present and test a new, special-relativistic formulation of Smoothed 
Particle Hydrodynamics (SPH). Our approach benefits from several improvements 
with respect to earlier relativistic SPH formulations. It is self-consistently 
derived from the Lagrangian of an ideal fluid and accounts for the terms that 
stem from non-constant smoothing lengths, usually called "grad-h terms".  In 
our approach,  we evolve the canonical momentum and the canonical energy per 
baryon and thus circumvent some of the problems that have plagued earlier 
formulations of relativistic SPH. We further use a much improved artificial 
viscosity prescription which uses the extreme local eigenvalues of the Euler 
equations and triggers selectively on a) shocks and b) velocity noise. The 
shock trigger accurately monitors the relative density slope and uses it to 
fine-tune the amount of  artificial viscosity that is applied. This procedure 
substantially sharpens shock fronts while still avoiding post-shock noise. 
If not triggered, the viscosity parameter of each particle decays to zero. 
None of these viscosity triggers is specific to special relativity, both could
also be applied in Newtonian SPH.\\ 
The performance of the new scheme is explored in a large variety of benchmark 
tests where it delivers excellent results. Generally, the grad-h terms deliver minor, 
though worthwhile, improvements. As expected for a Lagrangian method, it performs
close to perfect in supersonic advection tests, but also in strong relativistic 
shocks, usually considered a particular challenge for SPH, the method yields 
convincing results. For example, due to its perfect conservation properties, it
is  able to handle Lorentz-factors as large as $\gamma= 50 \; 000$ in the so-called
wall shock test. Moreover, we find convincing results in a rarely shown, but 
challenging test that involves so-called relativistic simple waves and also in 
multi-dimensional shock tube tests.  
\end{abstract}

\begin{keyword}
computational fluid dynamics \sep shocks \sep special relativity \sep smoothed particle hydrodynamics
\end{keyword}

\end{frontmatter}
\section{Introduction}
\label{sec_intro}
Special-relativistic hydrodynamics has important applications in the fields of heavy ion collisions
and in astrophysics. Astrophysical examples that involve highly relativistic motion include jets from 
Active Galactic Nuclei \cite{celotti01}, 
pulsar winds \cite{gallant94} or gamma-ray bursts  \cite{piran05} and often involve Lorentz factors 
substantially in excess of $\gamma= 10$. Analytical solutions are only known
for a small set of specific problems, for most relevant cases numerical approaches are required.
A robust special-relativistic scheme can be directly applied to problems of the above type and, in 
addition, it is a core ingredient for  general-relativistic codes, tools that open up the possibility 
to tackle a whole new class of interesting astrophysical phenomena.\\
In recent years, grid-based methods for special-relativistic hydrodynamics have seen a huge leap 
forward, and by now many highly accurate Eulerian schemes exist, see \cite{marti03} for a review. 
Most of these schemes are fine-tuned to solve 1D relativistic shock problems without oscillations and 
with sharp discontinuities. For many astrophysical problems, however, additional capabilities
such as the accurate advection of smooth flow features are required. In particular,
for some of the future applications that we have in mind a purely Lagrangian method possesses distinct 
advantages and this is why we focus here on a refined special-relativistic formulation of the Smoothed Particle 
Hydrodynamics (SPH) method.\\
SPH is a Lagrangian, purely mesh-free particle method.
Since its first formulations in an astrophysical context \cite{lucy77,gingold77} SPH has undergone 
a slew of technical improvements and it has  found its way into many other branches of computationally 
oriented areas of science. For reviews of the method see \cite{benz90a,monaghan92,monaghan05,rosswog09b}.
Being entirely Lagrangian, the method has obvious advantages in advection problems, on the other hand, 
strong shocks have traditionally posed serious challenges. Our aim is to devise a special-relativistic 
SPH formulation that, at least at decent resolution, yields accurate shock-results while keeping the other 
benefits of a Lagrangian scheme. This formulation is intended to become the condensation nucleus for a 
future, fixed-metric implementation of general relativistic SPH, the corresponding equations  can also be derived from a variational principle \cite{monaghan01,rosswog10a}.\\
The paper is organized as follows. In Section 2 we derive a SPH formulation consistently from
the Lagrangian of an ideal fluid and the first law of thermodynamics and we present the details of 
our treatment of artificial dissipation.
In Section 3 we investigate the performance of the new equation set in a number of 
special-relativistic benchmark tests. They are complemented by the tests shown in 
\cite{rosswog10d}. The main results will be summarized in Section 4.

 \section{Special-relativistic SPH with grad-h terms}
In the SPH discretization process, derivatives are expressed as 
sums over particle properties, weighted with the gradient of a smoothing kernel whose width is determined 
by the so-called smoothing length. If symmetrized appropriately, the SPH-discretized fluid equations 
conserve mass, energy, linear and angular momentum by construction. In early SPH formulations 
the derivatives of the kernel functions with respect to the smoothing length were assumed to vanish.
In practice, however, the smoothing lengths were still evolved to ensure a local adaptivity 
and this inconsistency lead to a violation of the conservation properties. How severe this violation is 
in practice, depends on the considered problem \cite{springel02,price04c,rosswog07c}.
This deficiency was first addressed by \cite{nelson94} and, more recently, by \cite{springel02} and 
\cite{monaghan02} who derived the SPH equations from a discretized fluid Lagrangian. The latter 
two approaches yielded correction factors for the kernel gradients, the so-called ``grad-h terms''.\\
In contrast to  earlier relativistic SPH formulations  \cite{kheyfets90,mann91,mann93,laguna93a,chow97,siegler00a} 
we derive our equation set  from a variational principle, similar to \cite{monaghan01},
but we also account for the kernel derivatives with respect to the smoothing length. For the flat-space metric tensor, 
$\eta_{\mu\nu}$, we use the signature (-,+,+,+), Latin indices run over 
(1,2,3), Greek ones run from 0 to 3 with the zero component being time. 
We apply the Einstein sum convention and use $c=1$ unless otherwise noted. 
With these conventions the four-velocity, $U^\mu=dx^\mu/d\tau$ is 
normalized to $U_\mu U^\mu=-1$.\\
In labeling the SPH particles we adhere to the following convention:
the particle of interest is always labeled $a$ and  neighbor particles,
e.g. those in the sum of Eq.~(\ref{eq:dens_summ_SR}), are usually denoted by $b$.
If some expression applies to both particles $a$ and $b$ of a previous expression,
we use the index $k$.

\subsection{The Lagrangian}
The Lagrangian of a perfect fluid can be written as \cite{fock64} 
\be
L_{\rm pf,sr}= - \int T^{\mu\nu} U_\mu U_\nu \; dV\label{eq:fluid_Lag_SRT},
\ee
where
\be
T^{\mu\nu}= (n[1 + u(n,s)] + P)  U^\mu U^\nu + P \eta^{\mu\nu}
\ee
denotes the energy momentum tensor, $n$ is the baryon number density in the 
local fluid rest frame, $u$ is the thermal energy per baryon, $s$ the specific 
entropy and $P$ the pressure. All these quantities are measured in the local 
rest frame of each fluid element, energies are measured in units of the baryon
rest mass energy\footnote{The appropriate mass $m_0$ obviously
  depends on the ratio of neutrons to protons, i.e. on the nuclear composition
of the considered fluid.}, $m_0 c^2$.
By using the normalization of the four-velocity, the Lagrangian
simplifies to
\be
L_{\rm pf,sr}= - \int n(1+u) \; dV. \label{eq:fluid_Lag_SRT_simp}
\ee
In the general case, a fluid element moves with respect to the frame in which the computations
are  performed (``computing frame'', CF). Therefore,  the baryon number density  
in the CF, $N$, is related to the local fluid rest frame via
a Lorentz contraction
\be
N= \gamma n, \label{eq:N_vs_n}
\ee
where $\gamma$ is  the Lorentz factor of the fluid element as measured in the CF.
The simulation volume in the CF can be subdivided into volume elements
such that each element $b$ contains $\nu_b$ baryons
\be
\Delta V_b= \frac{\nu_b}{N_b}.
\ee
These volume elements are used in the SPH discretization process to approximate 
a quantity $f$ given at a set of discrete points (``particles'') labeled by $b$:
\be
f(\vec{r})= \sum_b f_b \frac{\nu_b}{N_b} W(|\vec{r}-\vec{r}_b|,h),\label{eq:SPH_discret}
\ee
where our notation does not distinguish between the approximated values (the
$f$ on the LHS) and the values at the particle positions ($f_b$ on the
RHS). The quantity $h$ is the smoothing length that characterizes the width
of the smoothing kernel $W$.
The discretization prescription, Eq.~(\ref{eq:SPH_discret}),
yields for the  baryon number density in the computing frame:
\be
N(\vec{r})= \sum_b \nu_b W(|\vec{r}-\vec{r}_b|,h).\label{eq:dens_summ_SR}
\ee
This equation takes over the role of the usual density summation of
non-relativistic SPH, $\rho (\vec{r})= \sum_b m_b
W(|\vec{r}-\vec{r}_b|,h)$. Since we keep the baryon numbers associated with each
SPH particle, $\nu_b$, fix, there is no need to evolve a continuity equation
and baryon number is conserved by construction. If desired, the continuity equation
can be solved though, see e.g. \cite{chow97}. 
The discretized fluid Lagrangian reads 
\be
L_{\rm SPH,sr}= - \sum_b \frac{\nu_b}{N_b} n_b [1+ u(n_b,s_b)],
\ee
or, by use of Eq.~(\ref{eq:N_vs_n})
\be
L_{\rm SPH,sr}= - \sum_b \frac{\nu_b}{\gamma_b} [1+ u(n_b,s_b)].
\ee

\subsection{The momentum equation}
The momentum evolution of a particle $a$ follows from the Euler-Lagrange equations
\be
\frac{d}{dt} \frac{\p L_{\rm SPH,sr}}{\p \vec{v}_a} - \frac{\p L_{\rm SPH,sr}}{\p \vec{v}_a}= 0.
\ee
In calculating the baryon number density of particle, $b$, we use $b$'s own smoothing length
\be
N_b= \sum_k \nu_k W(|\vec{r}_b-\vec{r}_k|,h_b)\label{eq:dens_summ_SR_N_b}
\ee
and adapt the smoothing length according to 
\be
h_b= \eta \left(\frac{\nu_b}{N_b}\right)^{1/D}\label{eq:h_N},
\ee
where $\eta$ is a suitably chosen numerical constant, usually chosen around 1.5,
and $D$ is the number of spatial dimensions. 
Hence, similar to the non-relativistic case \cite{springel02,monaghan02}, the density and the
smoothing length mutually depend on each other and an iteration is
required to obtain a self-consistent solution for both. The density gradient with 
respect to particle position $a$ is given by
\bea
\nabla_a N_b&=& \sum_k \nu_k \left(\frac{\p W_{bk}(h_b)}{\p r_{bk}} \frac{\p
    r_{bk}}{\p \vec{r}_a} + \frac{\p W_{bk}(h_b)}{\p h_b} \frac{\p h_b}{\p N_b}
  \frac{\p N_b}{\p \vec{r}_a} \right)\nonumber\\
&=& \frac{1}{\tilde{\Omega}_b} \sum_k \nu_k \nabla_b W_{bk}(h_b) (\delta_{ba}-
\delta_{ka}), 
\label{eq:nabla_a_N_b}
\eea 
where the ``grad-h'' correction factor
\be
\tilde{\Omega}_b\equiv 1-\frac{\p h_b}{\p N_b} \sum_k \nu_k \frac{\p W_{bk}(h_b)}{\p h_b}
\ee
was introduced. Similarly, the time derivative becomes
\bea
\frac{dN_a}{dt}&=& \sum_b \nu_b \left(\frac{\p W_{ab}(h_a)}{\p r_{ab}} \frac{d
    r_{ab}}{dt } + \frac{\p W_{ab}(h_a)}{\p h_a} \frac{\p h_a}{\p N_a}
  \frac{d N_a}{dt} \right)\nonumber\\
&=& \frac{1}{\tilde{\Omega}_a} \sum_b \nu_b \vec{v}_{ab} \nabla_aW_{ab}(h_a).
\label{eq:dN_dt}
\eea
The canonical momentum is given by
\bea
\vec{p}_a &\equiv& \frac{\partial L_{\rm SPH,sr}}{\partial\vec{v}_a}
= - \sum_b \nu_b \frac{\partial}{\partial \vec{v}_a} 
\left(\frac{1+ u(n_b,s_b)}{\gamma_b}\right)
= \nu_a \gamma_a \vec{v}_a \left(1+u_a+\frac{P_a}{n_a}\right),
\label{eq:can_mom}
\eea
where we have used the first law of thermodynamics,
\be
\left(\frac{\p u_b}{\p n_b}\right)_{s}= \frac{P_b}{n_b^2}
\label{eq:1st_law}
\ee
and the relation between the baryon number densities in the different frames, 
Eq.~(\ref{eq:N_vs_n}). The last term in brackets on the RHS of Eq.~(\ref{eq:can_mom}) 
is the enthalpy per baryon. As numerical variable, we evolve the relativistic 
canonical momentum per baryon,
\be
\vec{S}_a\equiv \gamma_a \vec{v}_a \left(1+u_a+\frac{P_a}{n_a}\right) \label{eq_Sa}.
\ee
To find its evolution equation $\p L/\p \vec{r}_a$ needs to be calculated.
By once more using the chain rule, the first law of thermodynamics, 
Eq.~(\ref{eq:1st_law}), Eq.~(\ref{eq:N_vs_n}), Eq.~(\ref{eq:nabla_a_N_b}) 
and $\nabla_b W_{ba} = - \nabla_a W_{ab}$, which follows from the choice of a radial kernel,
$W(\vec{r})= W(|\vec{r}|)$, one finds
\bea
\frac{\partial L_{\rm SPH,sr}}{\partial\vec{r}_a}
&=& - \sum_b \frac{\nu_b}{\gamma_b} \frac{\partial u_b}{\partial
  \vec{r}_a}
= - \sum_b
\frac{\nu_b}{\gamma_b^2} \frac{P_b}{n_b^2} \nabla_a N_b\nonumber\\
&=& -\nu_a \sum_b \nu_b 
\left(
\frac{P_a}{N_a^2 \tilde{\Omega}_a} \nabla_a W_{ab}(h_a) + 
\frac{P_b}{N_b^2 \tilde{\Omega}_b} \nabla_a W_{ab}(h_b)
\right),
\eea
so that our special-relativistic momentum equation reads
\be
\frac{d\vec{S}_a}{dt}= - \sum_b \nu_b \left(
\frac{P_a}{N_a^2 \tilde{\Omega}_a} \nabla_a W_{ab}(h_a) + 
\frac{P_b}{N_b^2 \tilde{\Omega}_b} \nabla_a W_{ab}(h_b)
\right).\label{eq:momentum_eq_no_diss}
\ee

\subsection{The energy equation}
We use the canonical energy to identify a suitable energy variable. We find 
\be
E \equiv \sum_a \frac{\partial L}{\partial \vec{v}_a} \cdot \vec{v}_a - L
= \sum_a \nu_a \left(\vec{v}_a \cdot \vec{S}_a +
  \frac{1+u_a}{\gamma_a}\right),
\ee
which can be transformed into
\be
E= \sum_a \nu_a \left[\gamma_a \left(1+u_a+\frac{P_a}{n_a}\right) -
  \frac{P_a}{N_a}\right].
\ee
As numerical energy variable we choose the canonical energy per baryon
\be
\epsilon_a \equiv \gamma_a \left(1+u_a+\frac{P_a}{n_a}\right) -
\frac{P_a}{N_a}= \vec{v}_a \cdot \vec{S}_a + \frac{1+u_a}{\gamma_a}\label{eq:SR:epsilon_a}.
\ee
By using Eq.~(\ref{eq:N_vs_n}) once more one finds 
\be
\frac{d}{dt} \left(\frac{1+u_a}{\gamma_a}\right)= \frac{P_a}{N_a^2} \frac{dN_a}{dt} - \vec{S}_a \cdot \frac{d\vec{v}_a}{dt}\label{eq:SR:h1}
\ee
and therefore 
\bea
\frac{d \epsilon_a}{dt} &=& \frac{d}{dt} \left\{\vec{v}_a \cdot
  \vec{S}_a + \frac{1+u_a}{\gamma_a} \right\} = \vec{v}_a \cdot
\frac{d\vec{S}_a}{dt} + \frac{P_a}{N_a^2} 
\frac{dN_a}{dt}.
\label{eq:deps_dt}
\eea
By inserting Eqs.~(\ref{eq:dN_dt}) and (\ref{eq:momentum_eq_no_diss}) into (\ref{eq:deps_dt}), 
the energy equation becomes
\be
\frac{d \epsilon_a}{dt} = - \sum_b \nu_b 
\left(
\frac{P_a \vec{v}_b}{N_a^2 \tilde{\Omega}_a} \cdot \nabla_a W_{ab}(h_a) + 
\frac{P_b \vec{v}_a}{N_b^2 \tilde{\Omega}_b} \cdot \nabla_a W_{ab}(h_b)
\right),
\label{eq:ener_eq_no_diss}
\ee 
similar to the non-relativistic case when the "thermokinetic energy" $\frac{1}{2}v^2+u$
is evolved, e.g. \cite{monaghan05}. Alternatively, one could evolve the specific entropy,
for a discussion see \cite{springel02,muir03}.

\subsection{Artificial dissipation}
Our main aim is an accurate description of an ideal fluid without dissipation. We do require
however {\em local} artificial dissipation to produce entropy at shocks to ensure the proper
jump conditions, very similar to what nature does on scales well below the numerical 
resolution scale. In that sense one can think of  both artificial viscosity and Riemann solvers
as a subgrid model for physical viscosity that would act on an unresolvable scale.
Riemann solvers can be successfully used in SPH \cite{inutsuka02,cha03}, but often one prefers
a shock treatment via artificial viscosity which does not require the 
restriction to an ideal gas and that avoids the explicit solution of the Riemann problem.
Guided by the successes of relativistic Riemann solvers \cite{marti91}, Monaghan has 
constructed a new form of artificial viscosity terms \cite{monaghan97}.  We start from this
form of dissipative terms, but augment it by a new form of signal velocity and two triggers
that indicated when to apply it.

\subsubsection{The form of the dissipative terms}
The dissipative terms used in this work are similar to the suggestion of Chow and 
Monaghan \cite{chow97}
\be
\left(\frac{d\vec{S}_a}{dt}\right)_{\rm diss}= - \sum_b \nu_b \Pi_{ab} 
\overline{\nabla_a W_{ab}} \quad {\rm with} \quad \Pi_{ab}= - 
\frac{K v_{\rm sig}}{\bar{N}_{ab}} (\vec{S}_a^\ast-\vec{S}_b^\ast) \cdot\hat{e}_{ab}
\label{eq:diss_mom}
\ee
and 
\be
\left(\frac{d\epsilon_a}{dt}\right)_{\rm diss}=  - \sum_b \nu_b \vec{\Omega}_{ab} \cdot
\overline{\nabla_a W_{ab}} \quad {\rm with} \quad \vec{\Omega}_{ab} = -
\frac{K v_{\rm sig}}{\bar{N}_{ab}} (\epsilon_a^\ast-\epsilon_b^\ast)\hat{e}_{ab}. 
\label{eq:diss_en}
\ee
Here $K$ is a numerical constant of order unity, $v_{\rm sig}$ an
appropriately chosen signal velocity, see below, $\bar{N}_{ab}= (N_a+N_b)/2$, and
\be
\hat{e}_{ab}= \frac{\vec{r}_a-\vec{r}_b}{|\vec{r}_a-\vec{r}_b|}
\ee
is the unit vector pointing from particle $b$ to particle $a$. For the symmetrized kernel gradient
we use  
\be
\overline{\nabla_a W_{ab}} = \frac{1}{2}\left[\nabla_a W_{ab}(h_a) +  \nabla_a W_{ab}(h_b) \right].
\ee
Note that in \cite{chow97} $\nabla_a W_{ab}(h_{ab})$ was used instead of our choice $\overline{\nabla_a W_{ab}}$,
in practice we find the differences between the two symmetrizations negligible.
The stars at the variables in Eqs.~(\ref{eq:diss_mom}) and (\ref{eq:diss_en}) indicate that 
in Eqs.~(\ref{eq_Sa}) and (\ref{eq:SR:epsilon_a}) the projected Lorentz factors
\be
\gamma_k^\ast= \frac{1}{\sqrt{1-(\vec{v}_k\cdot \hat{e}_{ab})^2}}
\ee
are used instead of the normal Lorentz factor. 
This projection onto the line connecting particle $a$ and $b$ has been chosen to guarantee 
that the viscous dissipation is positive definite \cite{chow97}.

\subsubsection{Signal velocity}
The signal velocity that enters the artificial dissipation terms is an
estimate for the speed of approach of a signal sent from particle $a$ 
to particle $b$. The idea is to have a physically sound estimate that does 
not require much computational effort. In \cite{monaghan97,chow97} the numerical
solution of test problems was found to be rather insensitive to the exact 
form of $v_{\rm sig}$. For our formulation, we use 
\be
v_{\rm sig,ab}= {\rm max}(\alpha_a,\alpha_b),\label{eq:vsig}
\ee
where
\be
\alpha_k^{\pm}= {\rm max}(0,\pm \lambda^\pm_k)
\ee
with $\lambda^\pm_k$ being the extreme local eigenvalues of the Euler equations, see e.g. \cite{marti03},
\be
\lambda^\pm_k= \frac{v_\parallel(1-c_{\rm s}^2) \pm c_{\rm s} \sqrt{(1-v^2)(1-v_\parallel^2 - v_\perp^2 c_{\rm s}^2)}}{1-v^2 c_{\rm s}^2}
\ee
and $c_{{\rm s},k}$ being the relativistic sound velocity of particle $k$. In 1 D, this simply reduces
to the usual velocity addition law, $\lambda^\pm_k= (v_k\pm c_{{\rm s},k})/(1\pm v_k c_{{\rm s},k}) $.
The results are not particularly sensitive to the exact form of the signal velocity, but
in experiments we find that
Eq.~(\ref{eq:vsig}) yields somewhat crisper shock fronts and less smeared contact 
discontinuities (for the same value of $K$) than the suggestions of \cite{chow97}.

\subsubsection{Controlling the amount of dissipation}
\label{sec:AV_controle}
To ensure artificial viscosity does not influence the flow away from shocks, we make the involved
viscosity parameters time dependent, a strategy suggested by \cite{morris97} and subsequently successfully applied in several approaches, e.g. \cite{rosswog00,dolag05,rosswog07c}.  The 
art consists in finding triggers that indicate in which specific portion of  the flow, or more accurately,
at which SPH particle artificial dissipation is needed. It is needed both at the shock front itself and in 
addition possibly in the post-shock region to damp velocity noise.
What complicates things further is that both effects may need different amounts of viscosity, so that choosing 
a large viscosity that is able to resolve a strong shock  maybe more than what is needed to damp post-shock noise.
We therefore aim for two independent triggers: one that indicates a shock and 
another that triggers on noise in the velocity field. We follow here the recent suggestion of \cite{cullen10} to 
jump immediately to the desired value of the viscosity parameter rather than including the triggers in a 
source term \cite{morris97,rosswog00} that leads to a continuous rise of the viscosity parameter (which in some 
situations was found to be too slow \cite{cullen10}).\\
For a particle $a$ we determine a "desired" value of the viscosity parameter, see Eqs.~(\ref{eq:diss_mom}) 
and (\ref{eq:diss_en}), due to a possible shock, $K_{a,\rm shock}$, and one due to the possible presence of 
velocity noise, $K_{a,\rm noise}$. The desired value is then
\be
     K_{a,\rm des}= {\rm max}(K_{a,\rm shock},K_{a,\rm noise}). 
\ee
If $K_{a,\rm des} > K_a(t)$, we instantaneously set $K_a= K_{a,\rm des}$, 
otherwise $K_a(t)$ smoothly decays according to
\be
\frac{dK_a}{dt}= - \frac{K_a(t)-K_{\rm min}}{\tau_a},
\label{eq:K_evol}
\ee
where 
\be
\tau_a= \frac{\chi h_a}{{\rm min}_b(v_{{\rm sig},ab})}
\ee
is the decay time. We use $\chi= 20$ for a relatively slow decay. In the earlier approaches 
\cite{morris97,rosswog00} the parameter $K_{\rm min}$ was held 
at a small but finite value to keep particles well-ordered. Since our scheme
reacts immediately on noise, we can set $K_{\rm min}=0$. Our experiments did
not show any difference between a zero and a small non-zero value.\\
Similar to \cite{balsara95,cullen10}, this scheme could be further augmented by applying "limiters"
that suppress viscosity in cases where it is not desirable,  but this is beyond the scope of the current paper.

\begin{figure}[htbp]
     \centerline{                  
                        \includegraphics[width=8.5cm]{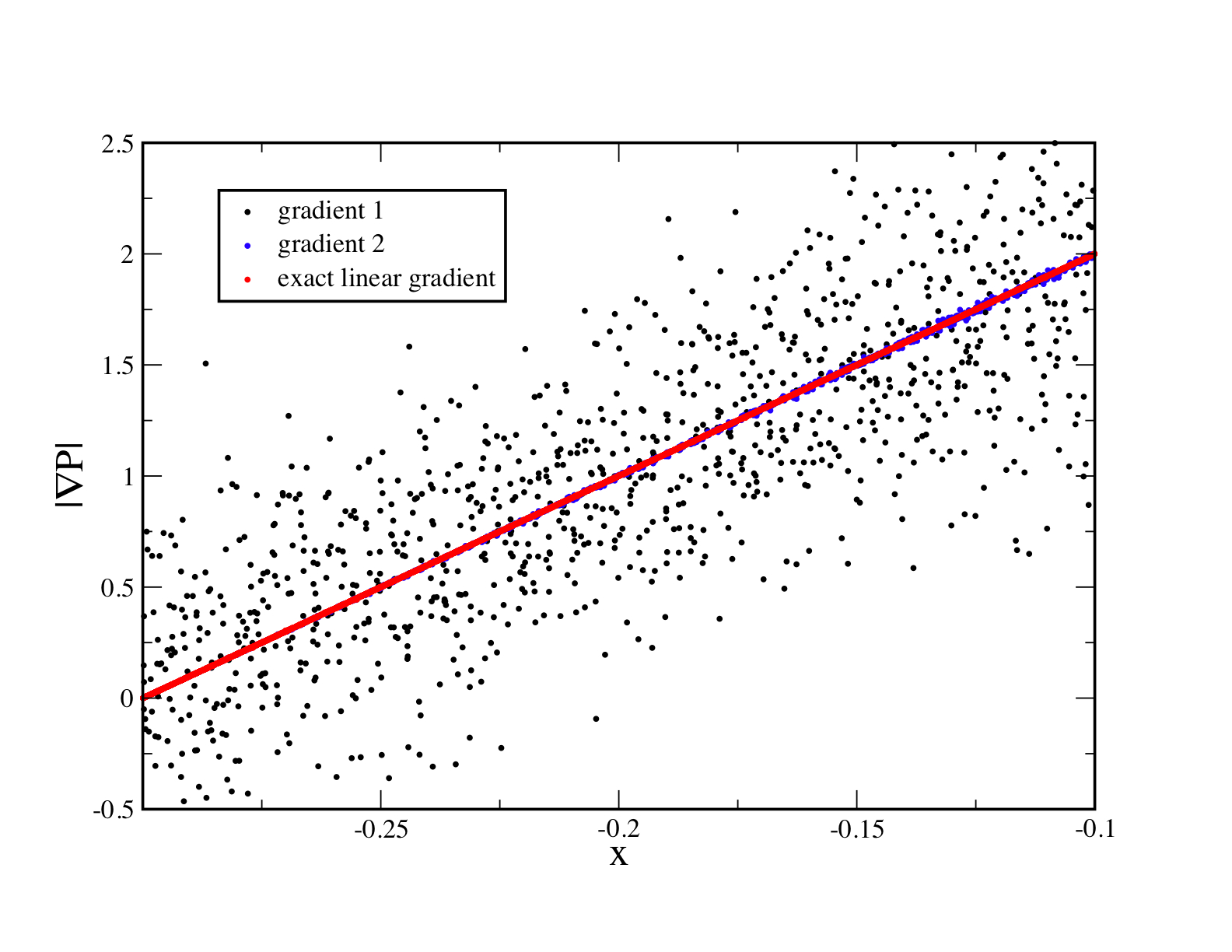}
                         \hspace*{-1.2cm}
                         \includegraphics[width=8.5cm]{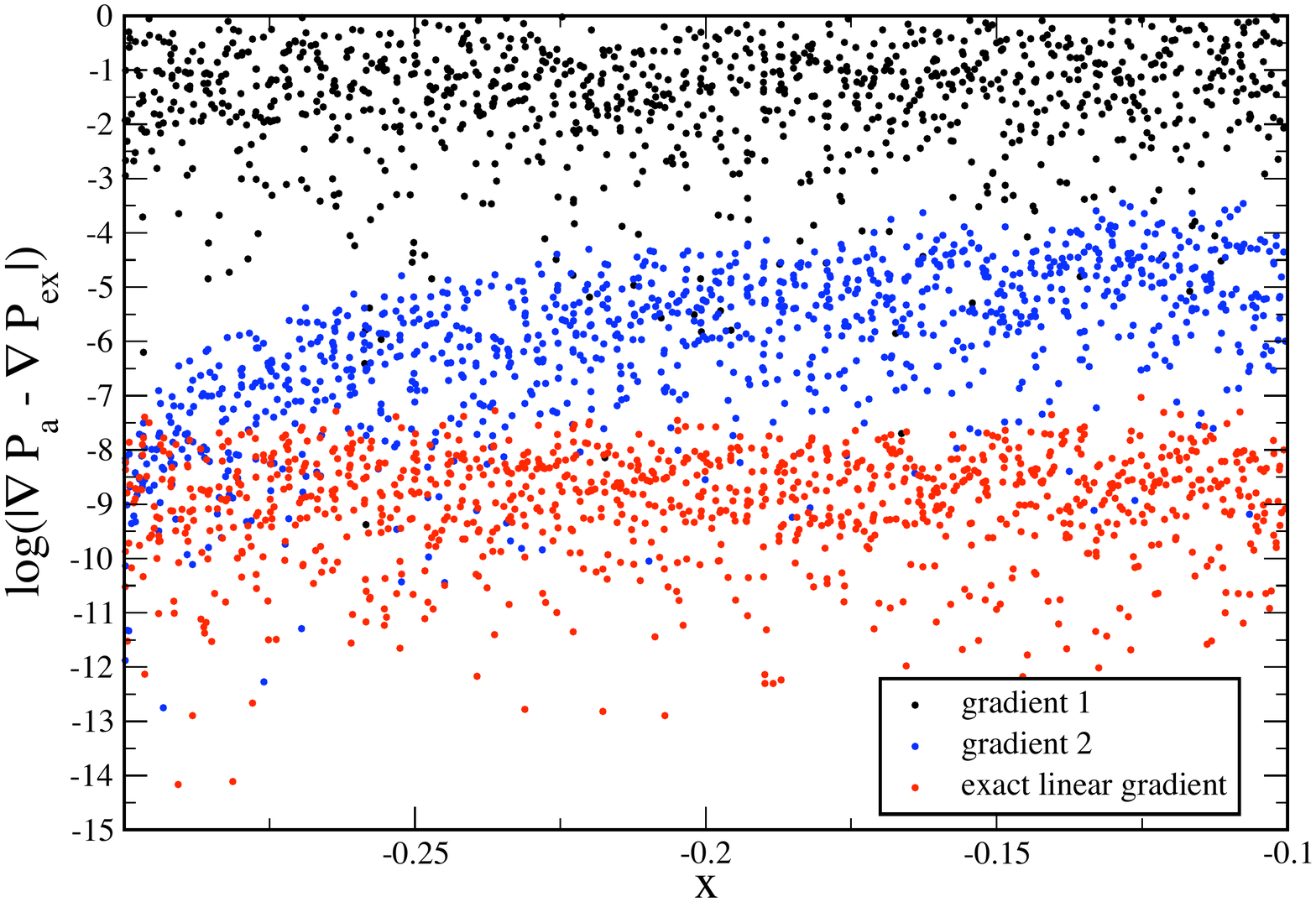}}                      
                         \caption{Accuracy of gradient estimates for a disordered particle distribution.
                         Left: gradient estimates for the different methods. Right: deviations from the exact result.}
   \label{fig:exact_lin_vs_SPH} 
\end{figure}

{\bf Shock trigger}\\
We use the local temporal change of the velocity divergence, $d (\nabla \cdot \vec{v})/dt$, to 
identify the emergence of shocks. This is different from earlier approaches \cite{morris97,rosswog00} 
which in their source terms trigger on $\nabla \cdot \vec{v}$ rather than on its temporal change. As 
already noted in the original paper \cite{morris97} such a scheme would also spuriously trigger on a 
constant slow compression with $\nabla \cdot \vec{v}$= const. Numerically, we calculate the divergence 
via "linearly exact derivatives", see below, and the temporal change by comparison with the last time 
step. Note that $d (\nabla \cdot \vec{v})/dt$ is not actually used to determine the desired dissipation
parameter value, it merely serves as indicator where to take action. This is different from \cite{cullen10} 
who used it to determine a desired viscosity parameter\footnote{In their work, Eq.(14), the viscosity parameter
is determined by a ratio of a physical and a resolution-dependent quantity. We suspect that at high resolution
this will produce too small a viscosity parameter.}. Instead,  we calculate the desired shock viscosity 
parameter via the relative change of the density $N$ across the kernel,
\be
A_{a,\rm shock}= \frac{|\nabla N|_a }{N_a}h_a,
\ee
where $\nabla N_a$ is again calculated via linearly exact derivatives, see below. Using this very local indicator of the density slope at the shock, we calculate the desired dissipation parameter  at the shock as
\be
     K_{a,\rm shock}=\left\{\begin{array}{ll}  K_{\rm max} \frac{A_{a,\rm shock}}{A_{\rm ref, shock} + A_{a,\rm shock}}& {\rm for} \; d (\nabla \cdot \vec{v})/dt > 0 \; {\rm and} \; \nabla \cdot \vec{v} < 0\\
         0, & {\rm else}\end{array}\right. .
\ee
In experiments, we find very crisp shocks for a reference value $A_{\rm ref, shock} = 0.9$ together with  $K_{\rm max}= 2$, 
but to be on the safe side we use $A_{\rm ref, shock} = 0.8$ in the following tests. This produced good results in both 1 and 2D.
Note also that the actually reached peak values of $K$ are usually substantially below $K_{\rm max}$, we will show examples  
in the context of the 2D shocks, see Sec.~\ref{sec:2D_shocks}.\\
We now want to address briefly the exact linear gradients. The derivative of a quantity $A$ 
with respect to coordinate $l$ at position $\vec{r}_a$ can be calculated as, see e.g. \cite{price04c,rosswog07c},
\bea
\left(\p_l A\right)_{\vec{r}= \vec{r}_a} = M^{lk} \tilde{A}_k = \bf{M} \cdot \vec{\tilde{A}}
\label{eq:exact_lin_grad}
\eea
with 
\be
M^{lk}= \left[\sum_b \nu_b (\vec{r}_b-\vec{r}_a)^l  \nabla_a^k W_{ab}\right]^{-1}
\ee
and
\be
\tilde{A}_k= \sum_b \nu_b (A_b-A_a) \nabla_a^k W_{ab} 
\ee
The matrix $M^{lk}$ corrects for effects from the particle distribution, so that linear functions 
are exactly reproduced even for an irregular distribution of particles.\\
To illustrate the accuracy of different gradient estimates, we perform a simple experiment.
We distribute SPH particles, once on a hexagonal lattice and once via the regularization sweeps 
described in Sec.~\ref{sec:2D_initial_particle_dist}, see Figs.~\ref{fig:grid_CP_distrib} (right)
and \ref{fig:particle_distributions} (middle), and assign them the same  density and baryon number 
and a pressure according to their positions $P(x)= 1 + (x_a - x_1)^2/(x_2-x_1)$ with $x_1= -0.3$ and $x_2= -0.1$.
Subsequently we calculate pressure gradients according to ("gradient 1")
\be
(\nabla P)_{a,1}= \sum_b \frac{\nu_b}{N_b} P_b \nabla_a W_{ab}(h_a),
\label{eq:sf_SPH_deriv}
\ee
or ("gradient 2")
\be
(\nabla P)_{a,2}= \sum_b \frac{\nu_b}{N_b} (P_b-P_a) \nabla_a W_{ab}(h_a),
\label{eq:sf_SPH_deriv1}
\ee
or according to the exact linear gradient. The second estimate is just an SPH estimate of $(\nabla P)_a - P_a \nabla(1)$,
see Eq.~(\ref{eq:SPH_discret}), which just subtracts the leading error term from Eq.~(\ref{eq:sf_SPH_deriv}).
For the case where the particles are located on the hexagonal lattice all estimates yield accurate results and lie on the same 
straight line as they should. For the irregular particle distribution, the first gradient estimate produces 
a substantial scatter (black) around the exact result, see left panel Fig.~\ref{fig:exact_lin_vs_SPH}. 
The second gradient estimate (blue) and the exact linear gradient result (red) are hard
to distinguish by eye, but the latter produces errors that are approximately four orders of magnitude smaller 
(Fig.~\ref{fig:exact_lin_vs_SPH}, right panel).

{\bf Noise trigger}\\
Our aim is to find an additional trigger that indicates "velocity noise" as it can appear behind
a shock. Such regions are characterized by some particles suffering expansion ($\divv >0$) while their 
neighbors feel a compression ($\divv <0$). Therefore, the ratio
\be
\frac{S_{1,a}}{S_{2,a}}\equiv \frac{\sum_b (\divv)_b}{\sum_b |\divv|_b} \label{eq:noise_trigg}
\ee
can deviate from $\pm 1$  in a noisy region since contributions of different sign are added up 
in $S_{1,a}$ and therefore such deviations can be used as a noise indicator. To remain very 
local and to avoid an unnecessary smearing of the shock front, our summation in 
Eq.~(\ref{eq:noise_trigg}) only runs over neighbors within $h_a$ (our kernel extends to $2 h_a$). We use
\be
A_{a, \rm noise}= \left| \frac{\tilde{S}_{1,a}}{S_{2,a}} - 1\right|,
\ee
where the quantity
\be
\tilde{S}_{1,a}= \left\{\begin{array}{ll}  -S_{1,a} & {\rm if} \; (\nabla \cdot \vec{v})_a < 0 \\
         \quad S_{1,a} & {\rm else}\end{array}\right. 
\ee
should be zero without noise and our noise trigger becomes
\be
     K_{a,\rm noise}=\left\{\begin{array}{ll}  K_{\rm max} \frac{A_{a,\rm noise}}{A_{\rm ref, noise} + 
        A_{a,\rm noise}}& {\rm for} \; S_{2,a} > 0.001 c_{{\rm s}, a}/h_a\\
         0 & {\rm else}\end{array}\right. 
\ee
For the noise reference value we use $A_{\rm ref, noise}= 5$. The threshold for $S_{2,a}$ was introduced to 
avoid triggering on acceptably tiny fluctuations around zero. Note that our choice of dissipation 
parameters is on the ``low-viscosity side'' and sometimes can produce small, but in our opinion
acceptable, oscillations. This can be cured, of course, by applying more dissipation.\\
The quantities $K_{a,\rm shock}$ and $K_{a,\rm shock}$ are stored as an accurate indicators of whether
a particle is in a shock or a noisy region.

\subsection{Smoothing kernel}
\begin{figure}[htbp] 
   \centering
   \includegraphics[width=12cm]{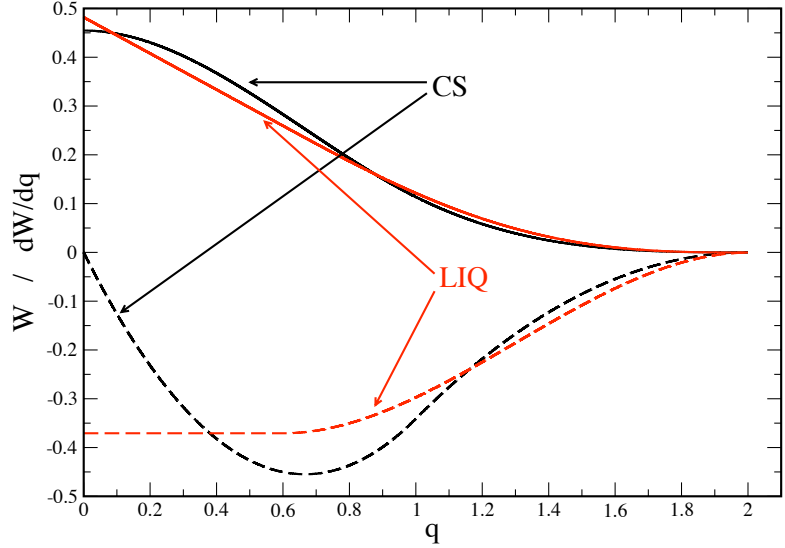} 
   \caption{Comparison of kernel $W$ (solid) and its derivative, $dW/dq$, (dashed) for the cubic spline (CS) and the linear-quartic (LIQ) kernel.}
   \label{fig:kernel_comparison}
\end{figure}

Traditionally, most SPH formulations use the cubic spline (CS) kernel suggested by \cite{monaghan85a},
\be
W_{\rm CS}(q)= \frac{N}{h^D}   \left\{
  \begin{array}{ l l l}
     1 - \frac{3}{2} q^2 + \frac{3}{4} q^3 \quad {\rm for \; } q \le 1\\
     \frac{1}{4} (2-q)^3 \quad  \quad \quad   {\rm for \; }    1 < q \le 2\\
     0   \quad  \quad  \quad \quad  \quad \quad \; \; {\rm else } 
       \end{array} \right.
  \ee
with $q=\frac{r}{h}$, $D$ the number of spatial dimensions and $N$ the normalization (2/3 in 1D, 
$10/7\pi$ in 2D and $1/\pi$ in 3D). It has been shown to yield good results over a large variety of 
test problems. This kernel, however, has the known shortcoming that its vanishing derivative at
$r=0$ allows particles to "pair" once they come close enough to each other, say in a shock. Often 
this has no dramatic effect, but it effectively reduces the resolution due to a poorer volume 
sampling  by the SPH particles. Recent investigations \cite{read10,valcke10} find that kernels 
that are centrally peaked perform better in Kelvin-Helmholtz instabilities since they enforce a 
more regular particle distribution across the contact discontinuity. We find very good results 
in 1D with the CS kernel, but in 2D tests we also explore the performance of the centrally peaked 
"Linear Quartic" (LIQ) kernel \cite{valcke10} in the form
\be
W_{\rm LIQ}(q)= \frac{N}{h^D}   \left\{
  \begin{array}{ l l l}
     F - u   \hspace*{4.5cm} {\rm for \; } u \le x_s\\
     A u^4 + B u^3 + C u^2 + D u + E  \hspace*{0.5cm} {\rm for \; }    u \le 1\\
     0   \hspace*{5.4cm} {\rm else } 
       \end{array} \right.\ee
with $x_s=0.3, A=-1.458, B=3.790, C=-2.624, D=-0.2915, E=0.5831$ and $F= 0.6500$ and $u= q/2$. The normalization constant $N$ is 2.962 in 2D and 3.947 in 3D. A comparison of both kernels and their derivatives is shown in Fig.~\ref{fig:kernel_comparison}. The results of our  2D shock test 8 favor the LIQ over the CS kernel.

\subsection{Conversion between primitive and numerical variables}
The new numerical variables, $N$, $\vec{S}$ and $\epsilon$, obtained from 
the integration process, need to be converted into the physical quantities $\gamma$, $u$,
$n$ and $v$. We follow the strategies of earlier special-relativistic approaches \cite{marti96,monaghan97}: all variables in the (polytropic) equation of state
\be
P= (\Gamma-1) n u 
\label{eq:polytrope}
\ee
are expressed as a function of the updated numerical variables and the pressure 
itself. The resulting equation is solved numerically for the new pressure which
is subsequently used to recover the physical variables.
From Eq.~(\ref{eq_Sa}) and (\ref{eq:SR:epsilon_a}) one finds
\be
\vec{v}= \frac{\vec{S}}{\epsilon+P/N}
\label{eq:v_of_S_eps}
\ee
and thus 
\be
\gamma= \frac{1}{\sqrt{1-S^2/(\epsilon+P/N)^2}}.
\label{eq:gamma_of_P}
\ee
Using Eq.~(\ref{eq:v_of_S_eps}) and Eq.~(\ref{eq:N_vs_n}) one can express the
specific energy as
\be
u= \frac{\epsilon}{\gamma} + \frac{P}{\gamma N} (1-\gamma^2)-1. 
\label{eq:u_of_eps}
\ee
With aid of Eqs.~(\ref{eq:N_vs_n}) and (\ref{eq:u_of_eps})
Eq.~(\ref{eq:polytrope}) can be solved for the new pressure $P$ 
that corresponds to the new values of the integrated numerical variables. Once $P$ is
known, the Lorentz factor can be calculated from Eq.~(\ref{eq:gamma_of_P}),
the specific energy from Eq.~(\ref{eq:u_of_eps}) and the velocity from
Eq.~(\ref{eq:v_of_S_eps}).

\subsection{Time integration}
We use the optimal third-order TVD algorithm to integrate the system of ordinary differential
equations of the form $\d\vec{u}/dt= f(\vec{u})$. The solution is advanced by one time step $\Delta t$ to
time $t^{n+1}$ according to
\bea
\vec{u}^{(1)}   &=& \vec{u}^n + \Delta t f(\vec{u}^n)\\
\vec{u}^{(2)}   &=& \frac{3}{4}\vec{u}^n + \frac{1}{4} \vec{u}^{(1)} + \frac{1}{4} \Delta t \; f\left(\vec{u}^{(1)}\right)\\
\vec{u}^{n+1} &=& \frac{1}{3}  \vec{u}^n + \frac{2}{3} \vec{u}^{(2)} + \frac{2}{3} \Delta t \; f\left(\vec{u}^{(2)}\right)
\eea
and we choose a simple error estimate to control the time step. Since the computing frame number density plays a central role
in the discretization process we use it to measure the error growth rate:
\bea
\epsilon \approx \frac{{\rm max}_a \left( \frac{ \left| N_a^{n+1} - N_{a, RK2}^{n+1}\right|}{N_a^{n+1}}\right)}{\Delta t},
\eea
where $N_{a, RK2}^{n+1}$ is the density estimate at $t^{n+1}$ obtained after a second-order Runge-Kutta step. The new time step is
then chosen as $\Delta t^{\rm new}= s  (\epsilon_{\rm tol}/\epsilon)^{1/2} \; \Delta t^{\rm old}$. For the ``safety factor'' we use
$s= 0.8$ and for the tolerable error growth rate $\epsilon_{\rm tol}= 5 \times 10^{-4}$. We use this rather conservative 
time step choice in the tests presented below. We find, however, comparable results with a simpler time step choice similar to \cite{chow97},
where the time step, $\Delta t= {\rm min}_a  \; \Delta t_a$, is determined by the momentum change according to $\Delta t_{a}= 0.3 \sqrt{h_a/|d\vec{S}_a/dt|}$.
In our experiments energy and momentum are conserved to about one part in $10^{14}$.

\subsection{Reference formulation}
The performance of the new equation set is compared to the formulation of \cite{chow97} which produces
the best shock test results of all published SPH formulations that we are aware of\footnote{Note that \cite{chow97}
obtain their density estimate from integration rather than by summation.}
\bea
N_a&=& \sum_b \nu_b W_{ab}(h_{ab})\\
\frac{d \vec{S}_a}{dt}&=& - \sum_b \nu_b \left(\frac{P_a}{N_a^2} + \frac{P_b}{N_b^2} + \Pi_{ab,{\rm CM}}\right) \nabla_a W_{ab}(h_{ab})\\
\frac{d \epsilon_a}{dt} &=& - \sum_b \nu_b \left(\frac{P_a \vec{v}_b}{N_a^2} + \frac{P_b \vec{v}_a}{N_b^2} + \vec{\Omega}_{ab,{\rm CM}}\right) \cdot \nabla_a W_{ab}(h_{ab}),
\eea
where $h_{ab}= (h_a+h_b)/2$ and
\bea
\Pi_{ab,{\rm CM}} &=& \frac{K v_{\rm sig}}{N_{ab}} (\vec{S}_a^\ast-\vec{S}_b^\ast)\cdot\hat{e}_{ab} \quad {\rm and} \quad
\vec{\Omega}_{ab,{\rm CM}} = \frac{K v_{\rm sig}}{N_{ab}}(\epsilon_a^\ast-\epsilon_b^\ast)\hat{e}_{ab},
\eea
where they use a fixed value $K=0.5$ for their 1D tests. For the following comparisons we use their 
first suggestion for the signal velocity
\be
v_{\rm sig, CM}= \frac{c_a + |v^\ast_{ab}|}{1 + c_a |v^\ast_{ab}|} + \frac{c_b + |v^\ast_{ab}|}{1 + c_b |v^\ast_{ab}|} + |v^\ast_{ab}|, 
\quad {\rm where} \quad
v^\ast_{ab}= -\vec{v}_{ab}\cdot \hat{e}_{ab},
\ee
their suggested alternative yields very similar results \cite{chow97}.

\section{Numerical results}
\label{sec:setup}
We use equal mass particles in our tests, so that  the density information
is encoded in the particle separation. Since SPH smoothes "discontinuities" over a 
few resolution lengths, we consider it consistent with the spirit of the method
to start a simulation from initial conditions that are smooth enough to be properly
resolved by the method. Throughout the test bench, we approximate discontinuities in the initial 
conditions of a function $f$ via Fermi-functions
\be
f(x)= \frac{f_{\rm L}-f_{\rm R}}{1+\exp(\frac{x-x_{\rm S}}{\Delta x})} + f_{\rm R},
\label{eq:fermi}
\ee
where $f_{\rm L}$ and $f_{\rm R}$ are the values to the left and right of the discontinuity 
located at $x_S$ and $\Delta x$ is the characteristic transition length. We use half of 
the average of the left and right interparticle separation for $\Delta x$. 
The issue of smoothed initial conditions is more a matter of taste than of technical
requirement. A comparison between simulations with $\Delta x$ as described and $\Delta x= 0$
shows only minor differences, see below.\\
Unless otherwise noted, about 3000 particles are shown and a polytropic equation of state 
with an adiabatic exponent specific to each test is used.

\subsection{Tests in 1D}
\subsubsection{Test 1: "standard" relativistic shock tube}
\begin{figure}[htbp] 
   \centerline{
   \includegraphics[width=3.5in]{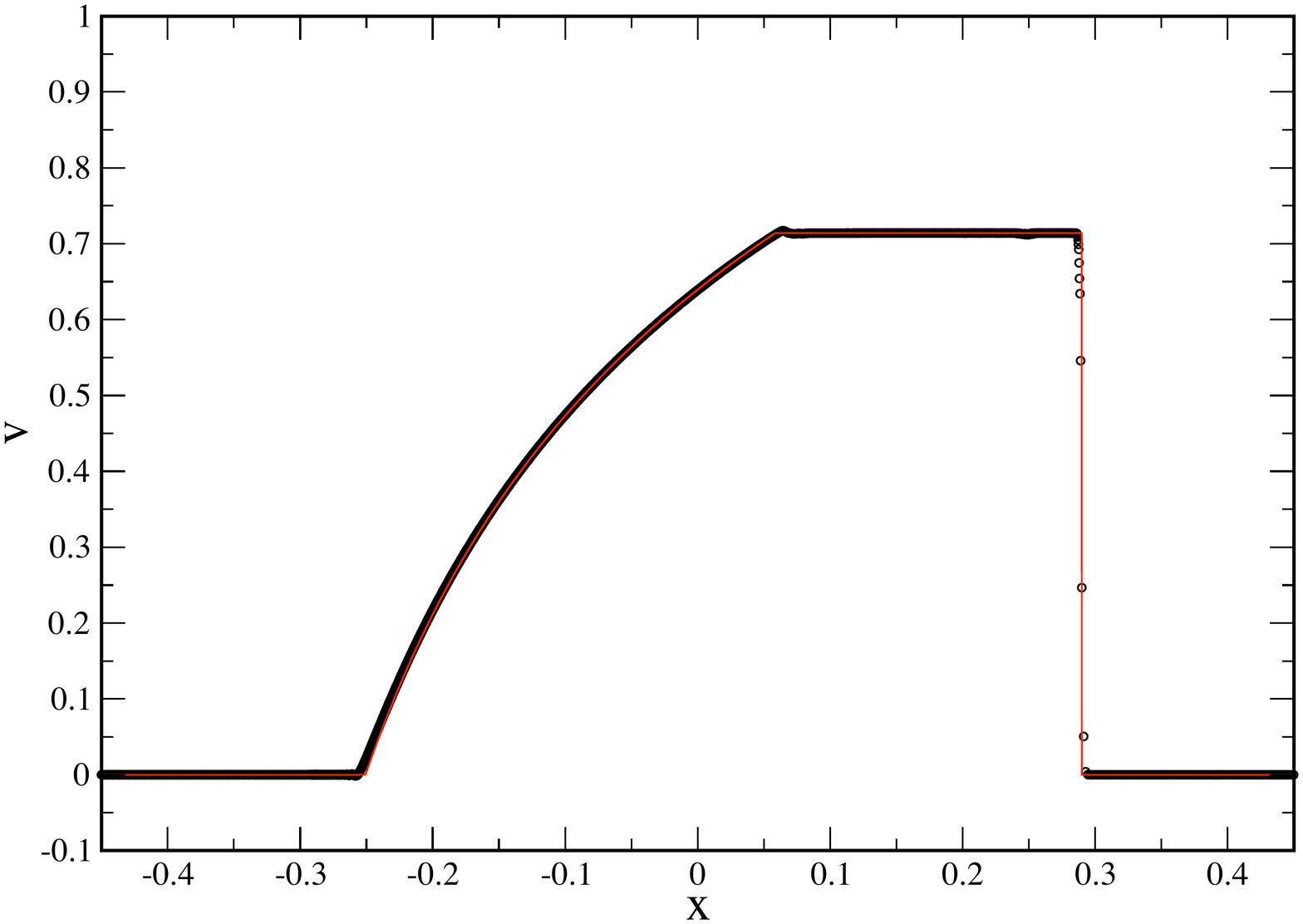}  
   \hspace*{-1.1cm}
   \includegraphics[width=3.5in]{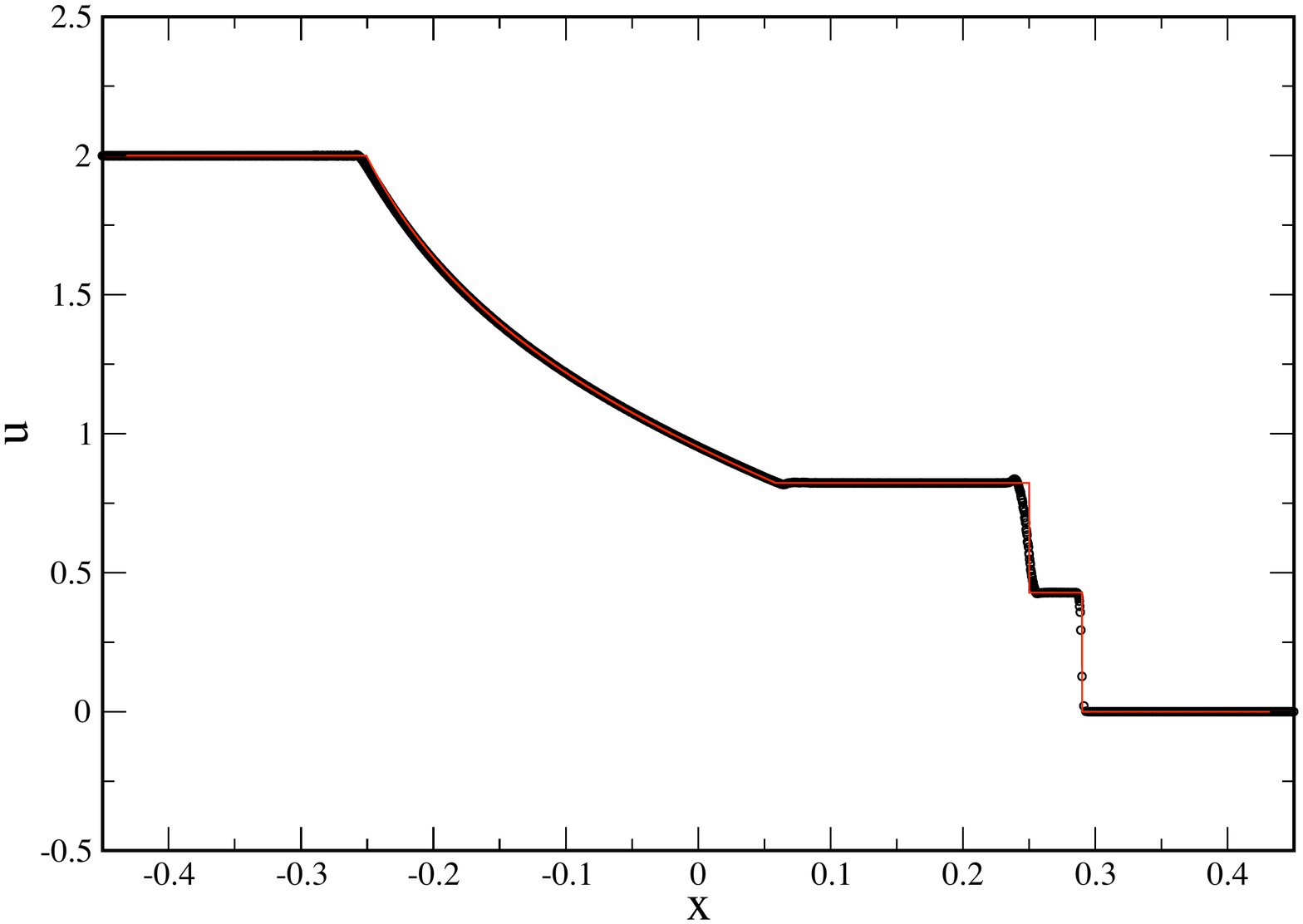} }
   \vspace*{-0.5cm}
   \centerline{
   \includegraphics[width=3.5in]{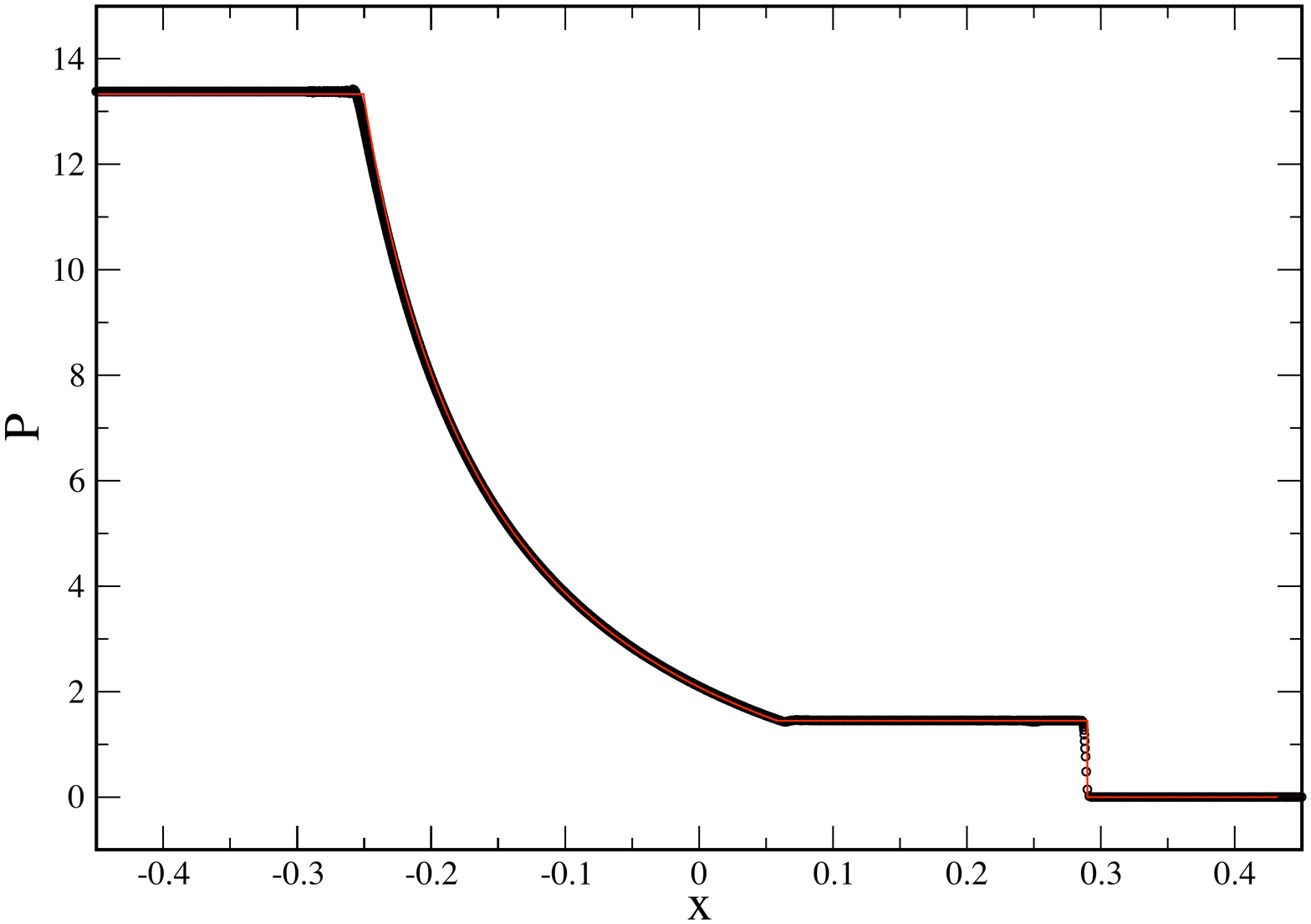}  
   \hspace*{-1.1cm}
   \includegraphics[width=3.5in]{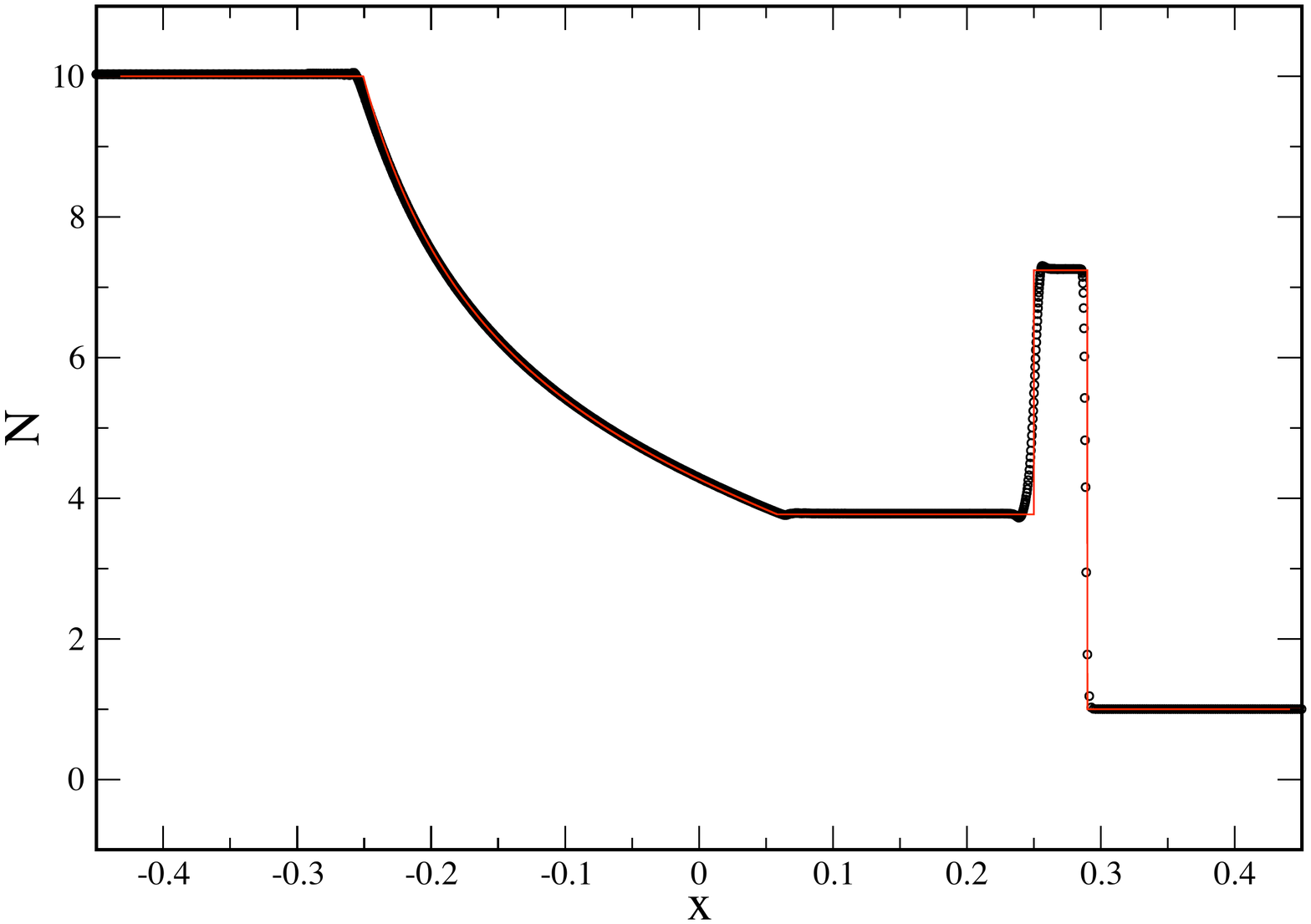} } 
   \caption{``Standard'' relativistic shock tube \cite{marti03} at t= 0.35: velocity (in units of the speed of light; 
   upper left), thermal energy (upper right), pressure (lower left) and computing frame number density (lower right).
   The SPH solution is shown as black circles, the exact solution as the red line.}
   \label{fig:Marti_Mueller_ST_I}
\end{figure}
\begin{figure}[htbp] 
   \centerline{\includegraphics[width=6in]{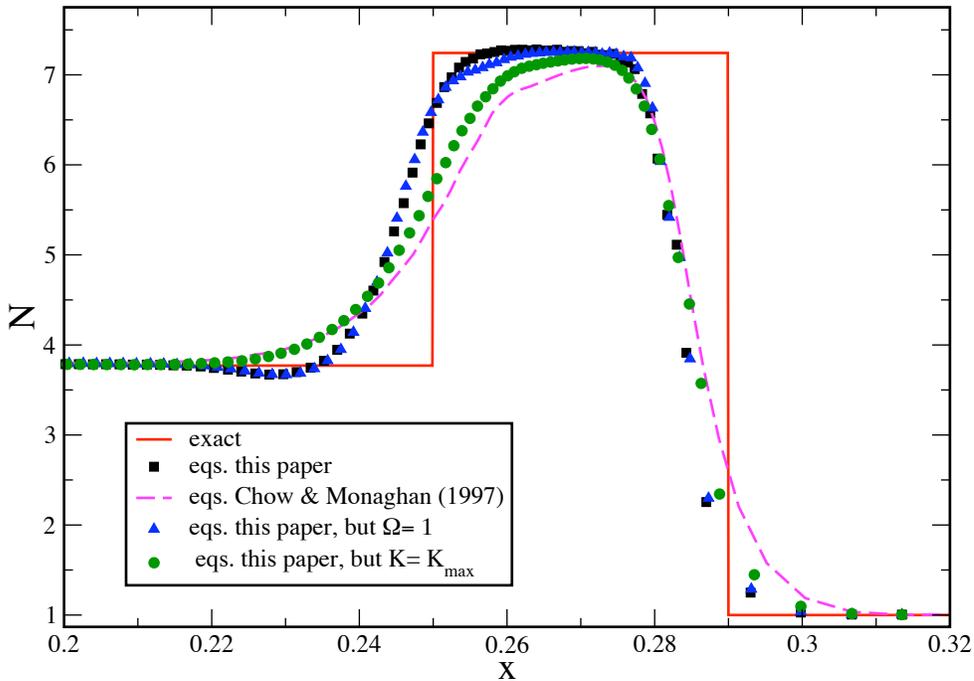}}
   \caption{Comparison of different equation sets, zoom into the density peak of a low resolution 
   simulation.}
   \label{fig:comparison_test1}
\end{figure}
\begin{figure}[htbp] 
   \centerline{\includegraphics[width=6in]{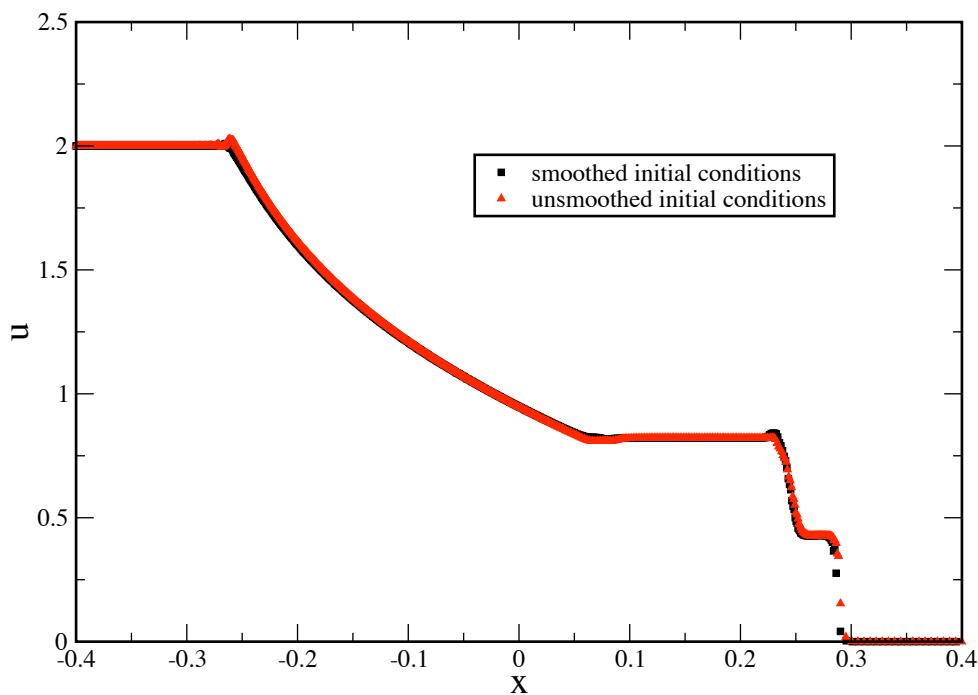}}
   \caption{Comparison between smoothed (black) and unsmoothed (red) initial conditions for the specific energy in test 1.}
   \label{fig:smoothed_vs_unsmoothed}
\end{figure}

This mildly relativistic shock tube ($\gamma_{\rm max} \approx 1.4$) has become a widespread benchmark
for relativistic hydrodynamics codes \cite{marti96,chow97,siegler00b,delZanna02,marti03}. It uses a polytropic
exponent of $\Gamma= 5/3$, vanishing initial velocities everywhere, the left state has a pressure
$P_L=40/3$ and a density $N_L=10$, while the right state is prepared with $P_R=10^{-6}$ and $N_R=1$.\\
The SPH result (circles, at t= 0.35) agrees excellently with the exact solution (solid line), 
see Fig.~\ref{fig:Marti_Mueller_ST_I}. Only the contact discontinuity at $x\approx 0.25$ is
somewhat smeared out. A striking difference to earlier SPH results \cite{laguna93a,siegler00a} is the absence
of any spike in $u$ and $P$ at the contact discontinuity. This is a result of the form of the dissipative terms, 
Eqs.~(\ref{eq:diss_mom}) and (\ref{eq:diss_en}).\\
To explore the dependence of the results on the various new elements we perform the following low-resolution 
(450 particles between -0.3 and 0.3) runs: i) use the new equation set, ii) the reference 
equation set of Chow and Monaghan \cite{chow97}
iii) the new equation set, but $\Omega=1$ to explore the importance of the ``grad-h''-terms 
and iv) the new equation set, but $K= K_{\rm max}= 0.5$ (the value chosen in\cite{chow97}) to explore the effect of the time-dependent viscosity parameters. With our parameters ($K_{\rm max}= 1.2,  A_{\rm ref, shock}= 0.8, A_{\rm ref, noise}= 5$) the dissipation parameter reaches 0.57, so slightly larger than 
the value chosen in \cite{chow97} to ensure a fair comparison. The results are displayed in Fig.~\ref{fig:comparison_test1}. All numerical parameters have exactly the same values in all cases.
Overall we find a good agreement between all the  equation sets. For a 
comparison we show the density $N$ since the shock-compressed shell is the most difficult structure
to capture and therefore shows the strongest deviations from the exact solution.
The "grad-h" terms improve the left edge of the rarefaction fan (not shown in the figure) and 
sharpen the left edge of the  shock-compressed shell (black square vs. blue triangle). The time-dependent viscosity parameters are substantially reduced behind the shock which allows the density
peak level to reach closer to the correct value (black squares vs. green circles).
The main difference between the suggested and the reference equation set  comes from the use of
the different signal velocity and the time-dependent dissipation parameters, the grad-h terms are 
only a minor, though welcome, improvement.\\
We also briefly compare smoothed vs unsmoothed initial conditions, see beginning of Section~\ref{sec:setup}.
To this end we use a low-resolution setup (about 700 particles) once smoothed using
$\Delta x= 0.5 (\Delta x_{\rm left}+\Delta x_{\rm right})$ in Eq.~(\ref{eq:fermi}) and once without smoothing, $\Delta x= 0$.
Here $\Delta x_{\rm left}$/$\Delta x_{\rm right}$ are the particle spacings on the left-/right-hand side. The result 
for the quantity that showed in earlier approaches the largest deviations from the exact result, the specific energy $u$,
are displayed in Fig.~\ref{fig:smoothed_vs_unsmoothed}. Overall, we find only minor differences.

\subsubsection{Test 2: strong blast}
The following test with initial conditions $(N,v,P)^{\rm L}=(1,0,1000)$ and $(N,v,P)^{\rm R}=(1,0,0.01)$
is a more relativistic variant of a shock tube and was first considered by
\cite{norman86}. It poses a severe challenge since relativistic effects compress the post-shock state into
a very thin and dense shell. The fluid in the shell moves at a velocity of $v= 0.96$ which corresponds to a 
Lorentz factor of $\gamma_{\rm shell}=3.6$, the shock front moves with a velocity of 0.986, i.e. $\gamma_{\rm shock}=6.0$. 
This test has become a standard benchmark for relativistic schemes 
\cite{norman86,dubal91,marti91,marquina92,marti96,falle96,wen97,chow97,donat98,delZanna02,anninos03,marti03}.
 \begin{figure}[htbp] 
   \hspace*{-0.4cm}\includegraphics[width=6in]{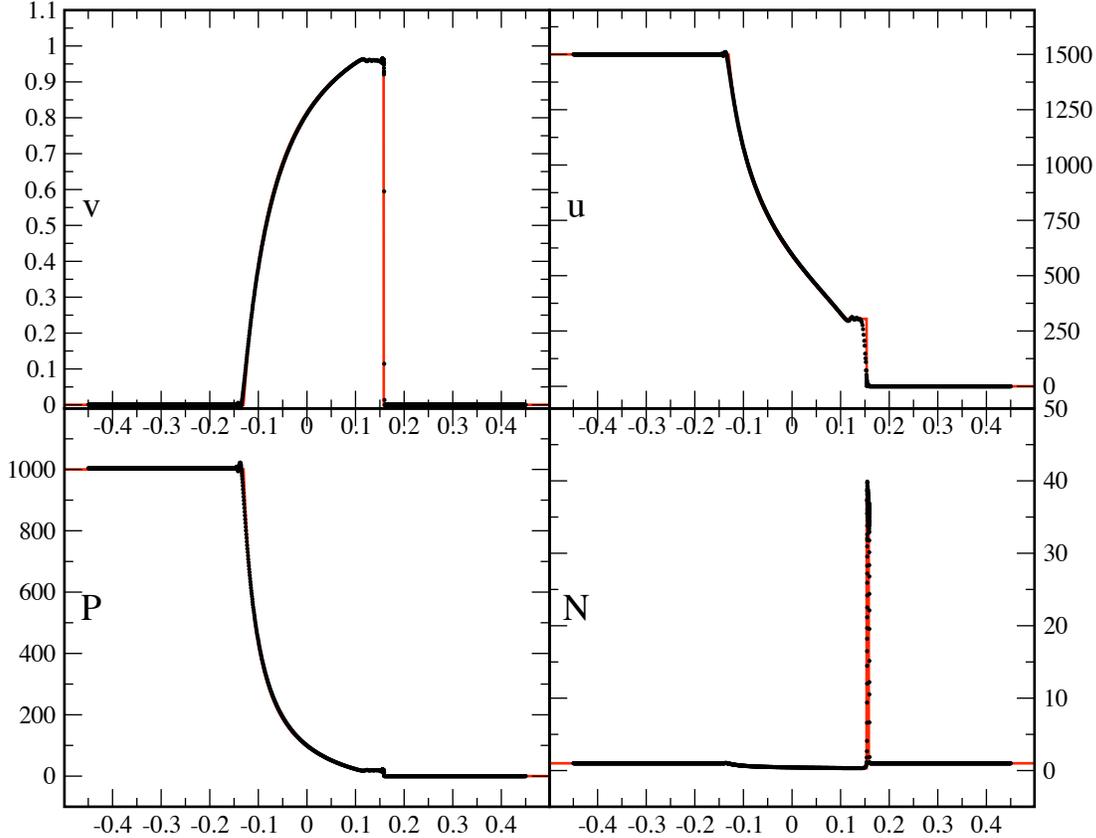}
      \caption{Strong blast \cite{marti03}: velocity (in units of the speed of light; upper left), thermal energy (upper right), pressure (lower left) and computing frame number density (lower right).
   The SPH solution is shown as black circles, the exact solution as the red line.}
   \label{fig:Marti_Mueller_ST_II}
\end{figure}
Overall, the numerical solution (shown at t=0.16, 1800 particles) agrees  well with the exact one, see Fig.~\ref{fig:Marti_Mueller_ST_II}. In particular, the intermediate states in velocity and pressure are well-captured. However, this difficult test is a severe challenge and the numerical solution
is not free of deficiencies. Somewhat large smearing in the internal energy occurs at $x\approx 0.15$, this is a result of using the maximum local eigenvalues rather than a proper spectral decomposition. Also, the numerical peak density value exceeds the exact one and the shock moves at a slightly too large velocity, effects that decrease with increasing numerical resolution. In comparison to \cite{chow97} both these artifacts are substantially reduced, but nevertheless still present.\\
We again perform a set of test runs (400 particles
between -0.5 and 0.5): i) use the new suggested equation set, ii) the reference equation set of Chow and 
Monaghan \cite{chow97} iii) the new suggested equation set, but $\Omega=1$ to explore the importance of 
the ``grad-h''-terms and iv) the new equation set, but $K= K_{\rm max}= 0.4$ (like in test ii) to explore 
the effect of the time-dependent viscosity parameters. The results are displayed in 
Fig.~\ref{fig:comparison_test2}.
 \begin{figure}[htbp] 
   \center{\includegraphics[width=5in]{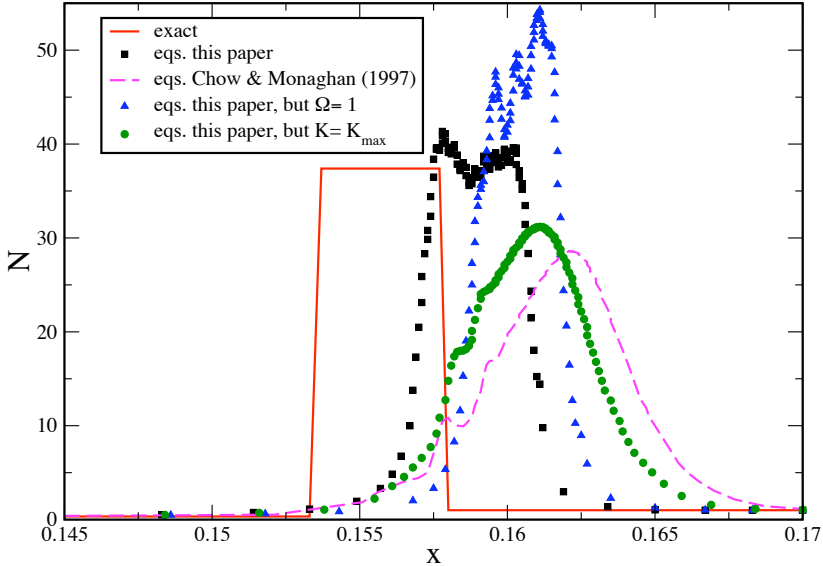}}
   \caption{Zoom into the density spike of low-resolution simulations with the different equation sets: 
    exact solution (solid, red), our new formulation (black squares), the new formulation, but 
$\Omega=1$ (blue triangles), the new formulation, but $K= K_{\rm max}$ (green circles) and the original formulation 
\cite{chow97} (dashed).}
   \label{fig:comparison_test2}
\end{figure}
At the given resolution, the new formulation (black squares) clearly performs best. This is the only test
where we see a clear improvement of the solution due to the grad-h terms (black squares vs. blue triangles).
Again, controlling the amount of dissipation has a major effect, the schemes with constant dissipation 
(green circles and dashed line) show the least satisfactory performance.

\subsubsection{Test 3: sinusoidally perturbed shock tube}
\begin{figure}[htbp] 
   \center{\includegraphics[width=5in]{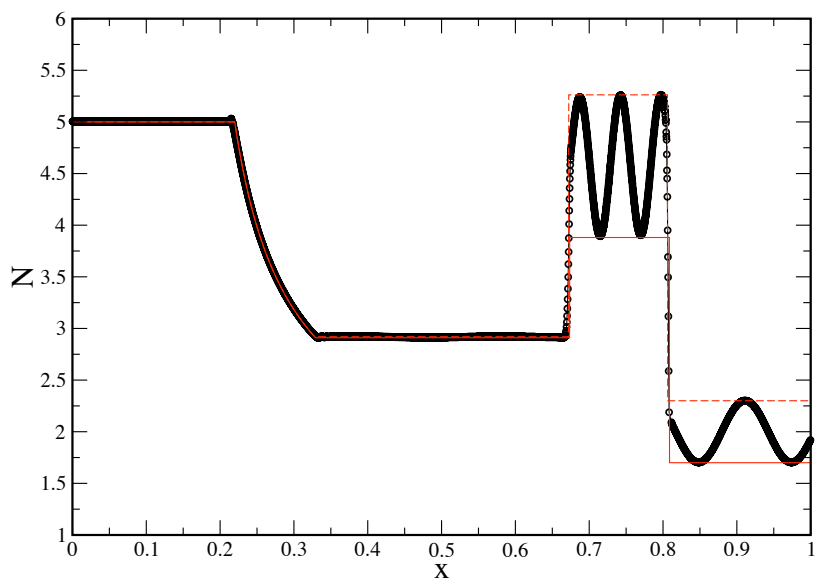}}
   \caption{Shock tube test where a relativistic shock propagates into a sinusoidally perturbed 
           medium \cite{dolezal95}. The SPH solution is shown as black circles, the red lines
           indicate the exact solutions for unperturbed right hand side densities of 1.7 (solid) and 2.3 (dashed).}
   \label{fig:perturbed_ST}
\end{figure}

Following  \cite{dolezal95}, we explore a shock tube test whose initial right density
state is sinusoidally perturbed:
\be
 (N,v,P)^{\rm L}=(5,0,50) \quad {\rm and} \quad (N,v,P)^{\rm R}=(2 + 0.3 \sin(50 x),0,5).
\ee
The main goal of this experiment is to test the ability to transport smooth structures across 
discontinuities. For this relatively mild shock only a moderate amount of dissipation is required, 
we use $K_{\rm max}= 0.2$. The numerical result at $t=0.35$ is displayed in Fig.~\ref{fig:perturbed_ST} together 
with the exact solutions of unperturbed shock tubes, once with a right hand side density value of 
2.3 (dashed red line) and once with 1.7 (solid red line). Note the slightly larger shock speed in the latter case 
($\gamma_{1.7} \approx 1.151$ vs. $\gamma_{2.3} \approx 1.149$). The numerical solution
accurately reaches the correct levels of the limiting solutions in the shocked shell.\\
We perform again test runs to explore the influence of the equation sets and parameters 
(400 particles between 0 and 1): i) the new equation set, ii) the reference equation set 
of Chow and Monaghan \cite{chow97} iii) the new equation set, but $\Omega=1$  and iv) the 
new equation set, but $K= K_{\rm max}$, where for the constant-$K$-cases (ii and iv) we 
use the maximum value from run i) (=0.2).The conclusions from this test set are similar to the 
previous tests: the effects of the ``grad-h-terms'' are visible, but small and the effects from the 
time-dependent viscosity parameters are substantially more important. These tendencies are found 
in all of the subsequent numerical experiments.

\subsubsection{Test 4: Ultra-relativistic wall shock}
\begin{figure}[htbp] 
   \centerline{
   \includegraphics[width=3.5in]{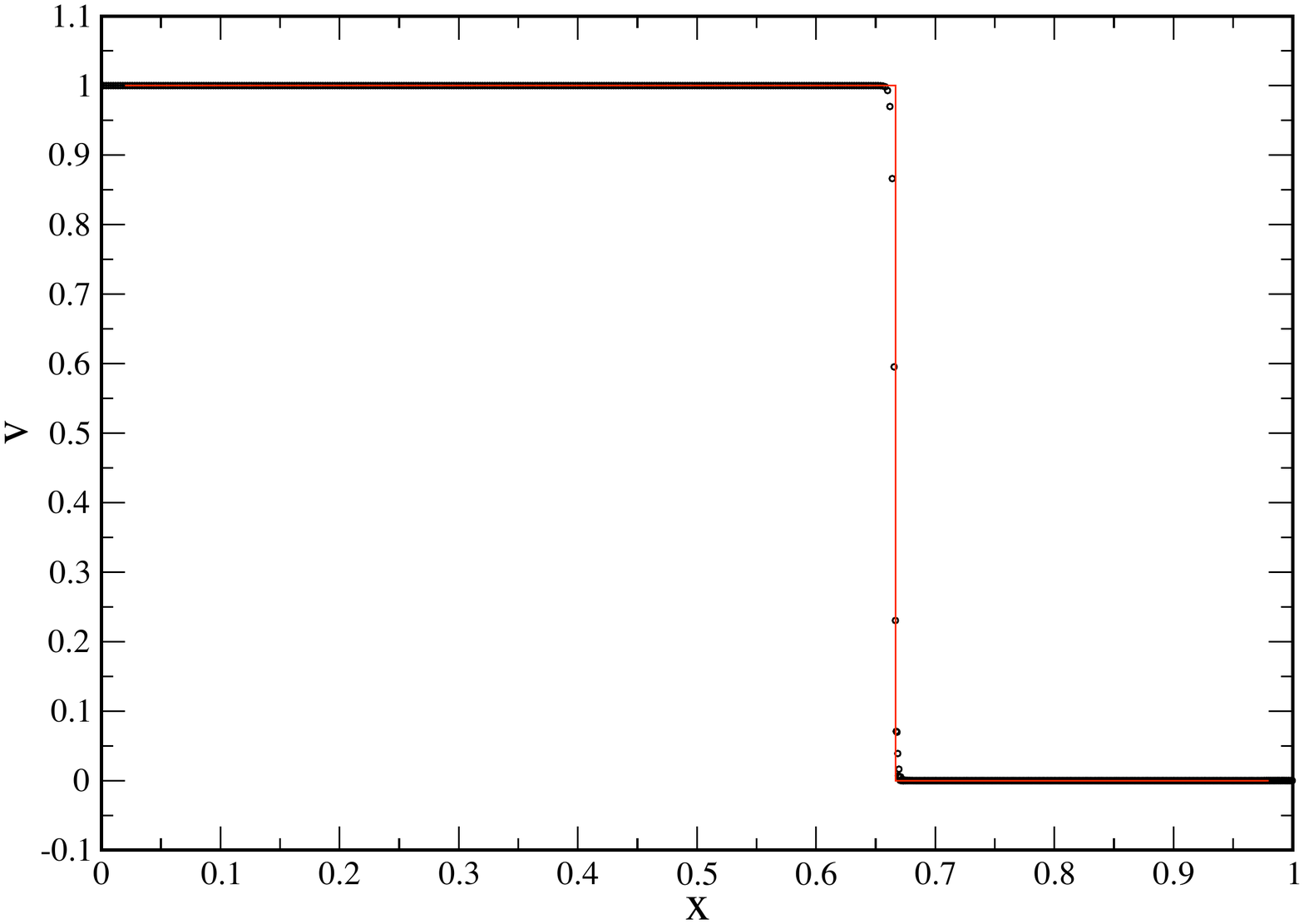}  
   \hspace*{-1.1cm}
   \includegraphics[width=3.5in]{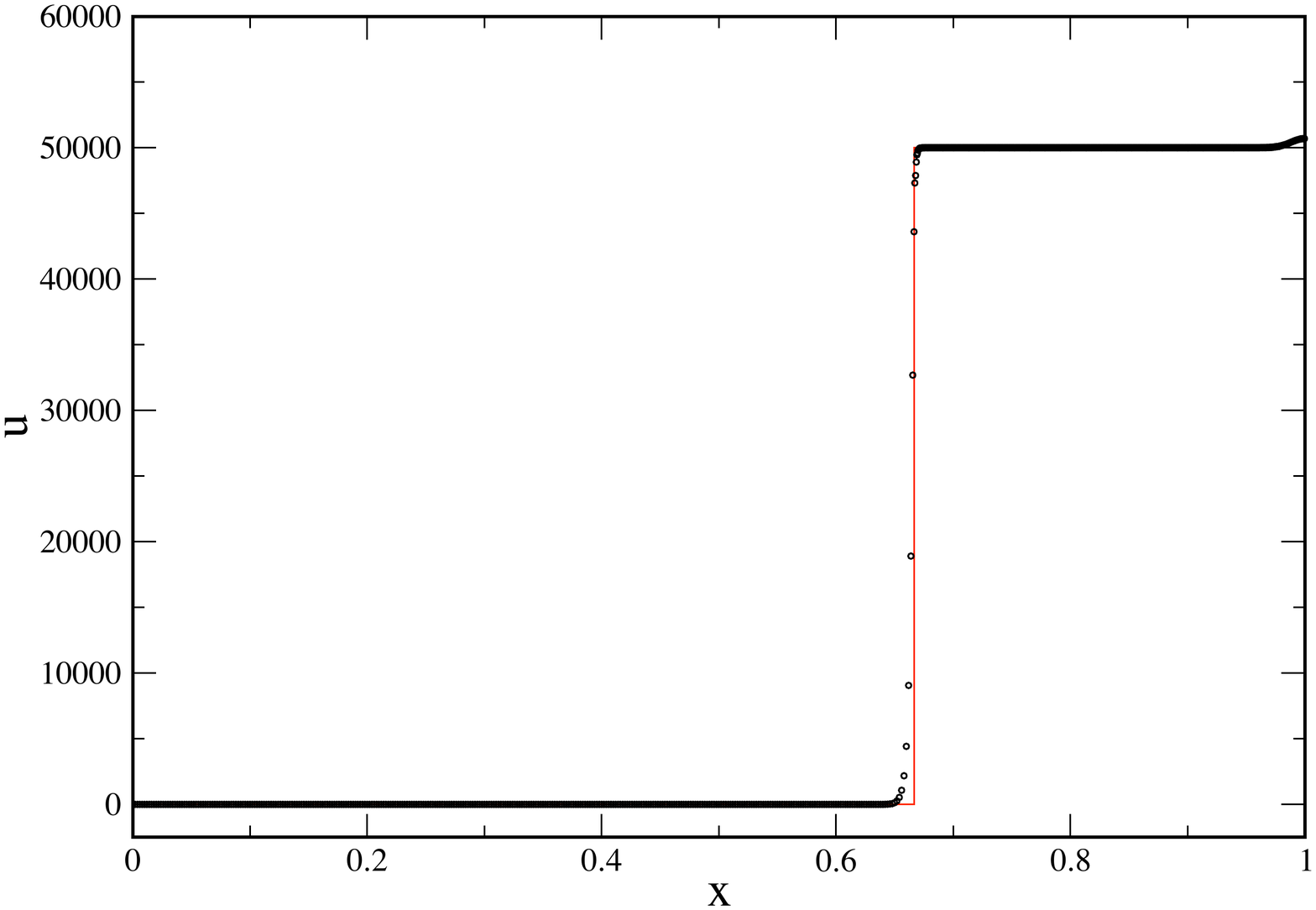} }
   \vspace*{-0.5cm}
   \centerline{
   \includegraphics[width=3.5in]{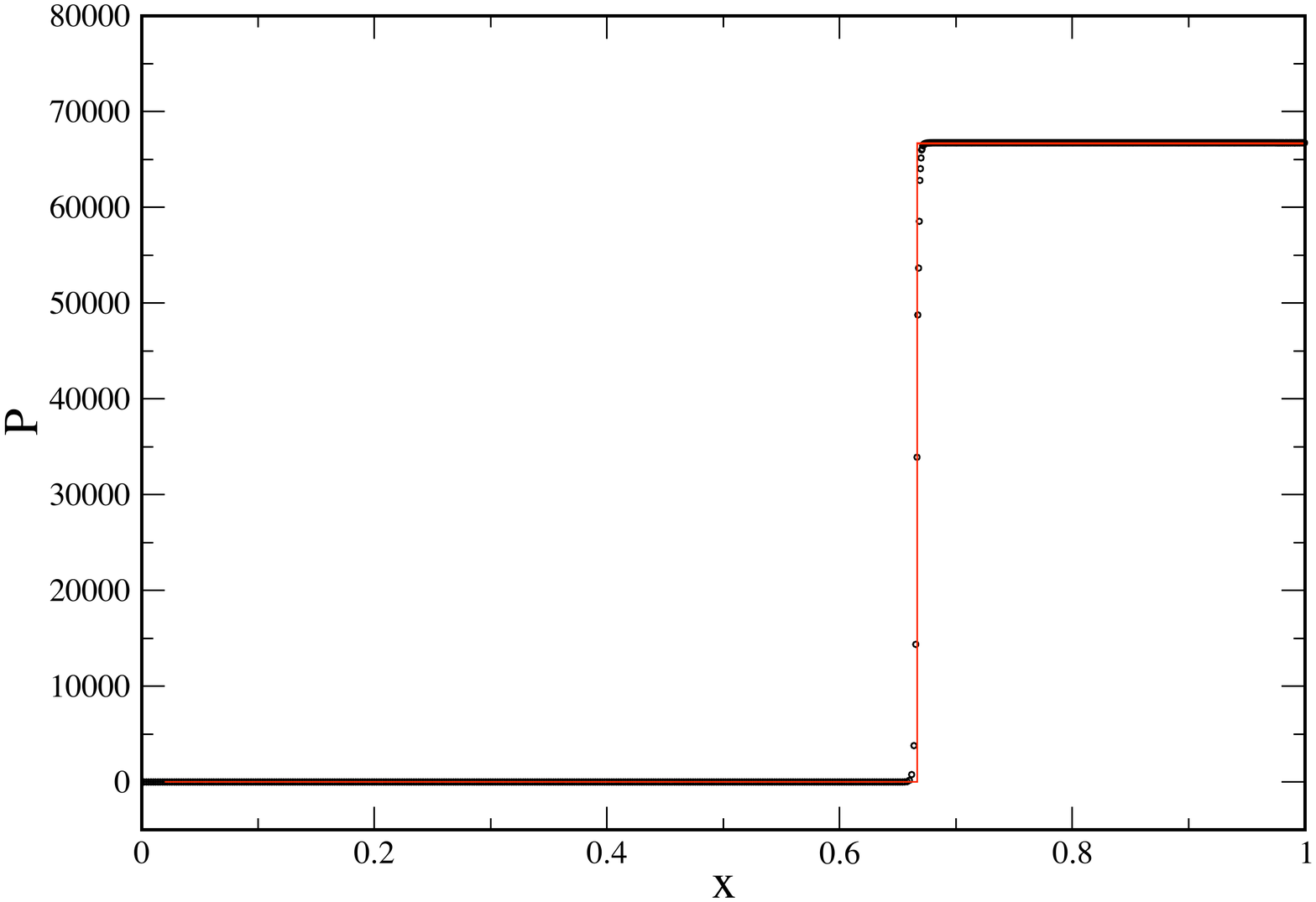}  
   \hspace*{-1.1cm}
   \includegraphics[width=3.5in]{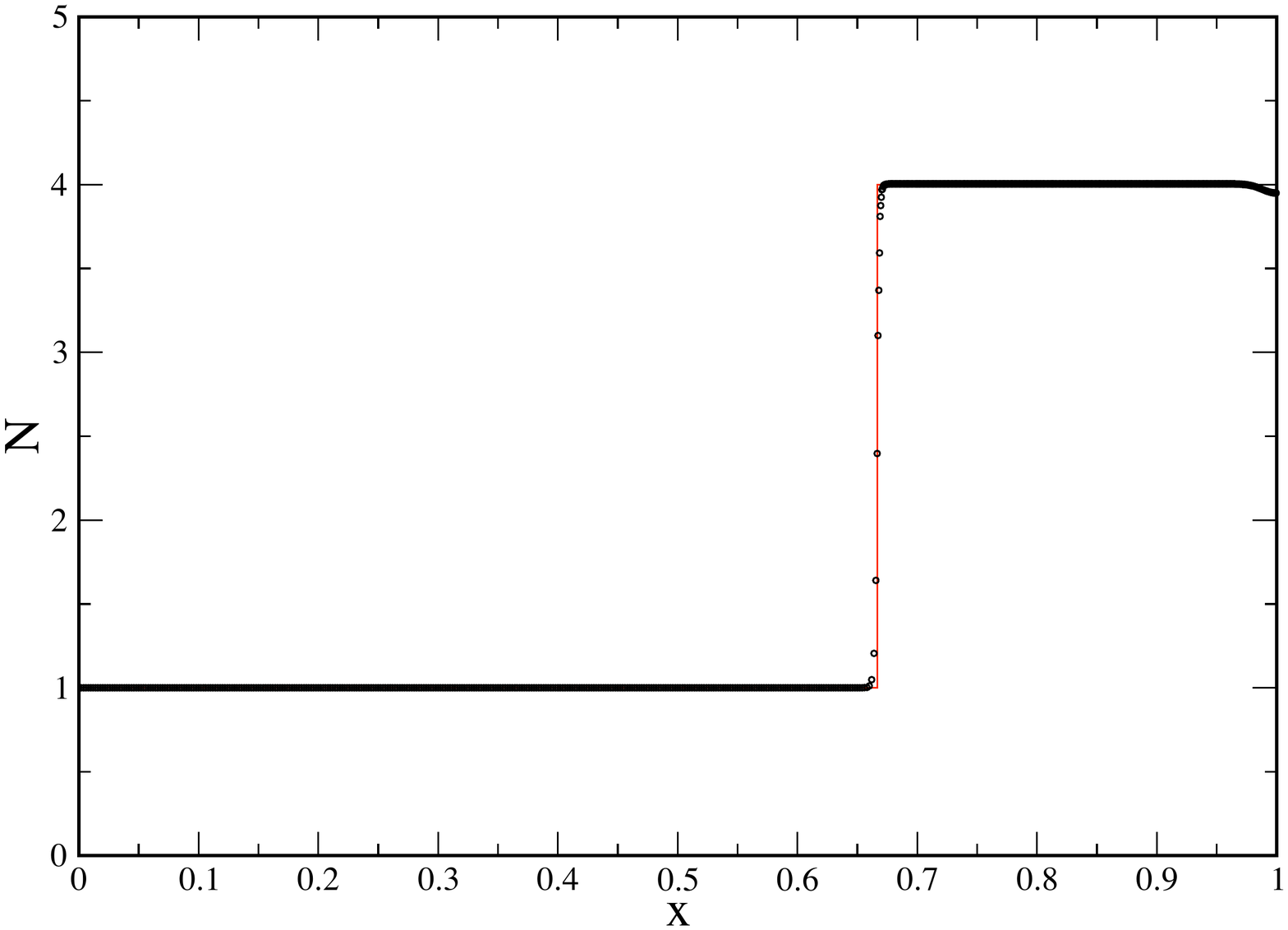} } 
   \caption{Ultra-relativistic wall shock test: an ultra-relativistic, cold fluid with $(N,v,u)= (1,0.9999999998,10^{-5}$)
            moves towards a reflecting wall at x=1. Note that the initial velocity corresponds to a Lorentz factor 
            as large as $\gamma= 50 000$. The SPH result is shown as black circles, the exact result in the ultra-relativistic limit
            as red lines.}
   \label{fig:wall_shock}
\end{figure}

In this test cold gas moves relativistically towards 
a wall. Upon hitting the wall, a shock front forms that travels upstream against the 
inflowing gas leaving behind a hot and dense post-shock region with zero velocity. 
In the ultra-relativistic limit ($v \rightarrow 1$) the shock travels at $v_{\rm shock}= - (\Gamma-1)$,
i.e. $v_{\rm shock}=1/3$ for the polytropic exponent $\Gamma=4/3$ that we use in this test.
The post-shock values of density, pressure and specific energy are $N_p= \Gamma/(\Gamma-1) N_i$, 
$P_p= \gamma \Gamma N_i$, $u_p= \gamma$, where $N_i$ is the initial density.\\
We model the reflecting wall as ``ghost'' particles streaming with opposite velocity from the 
right towards the wall located at $x=1$. For this extremely strong shock we use $K_{\rm max}= 1$. For the initial
gas velocity we use a value as high as $v= 0.9999999998$ corresponding to a Lorentz factor of 
50 000! We further use $N_i= 1$ and a specific energy of $u_i= 10^{-5}$.
The results of the numerical calculation (1000 particles) at t=1 are shown together with the ultra-relativistic 
limit values in Fig.~\ref{fig:wall_shock}.
The agreement between the numerical and the exact result is excellent, only the density and the specific energy
show minor deviations from the exact solution (maximum error in density 1.2 \%, in specific energy 1.1\%) as a 
result of so-called ``wall-heating'' \cite{norman86} near the boundary at x=1.

\subsubsection{Test 5: Relativistic advection of a sine wave}
\begin{figure}[htbp] 
   \centerline{
   \includegraphics[width=5in]{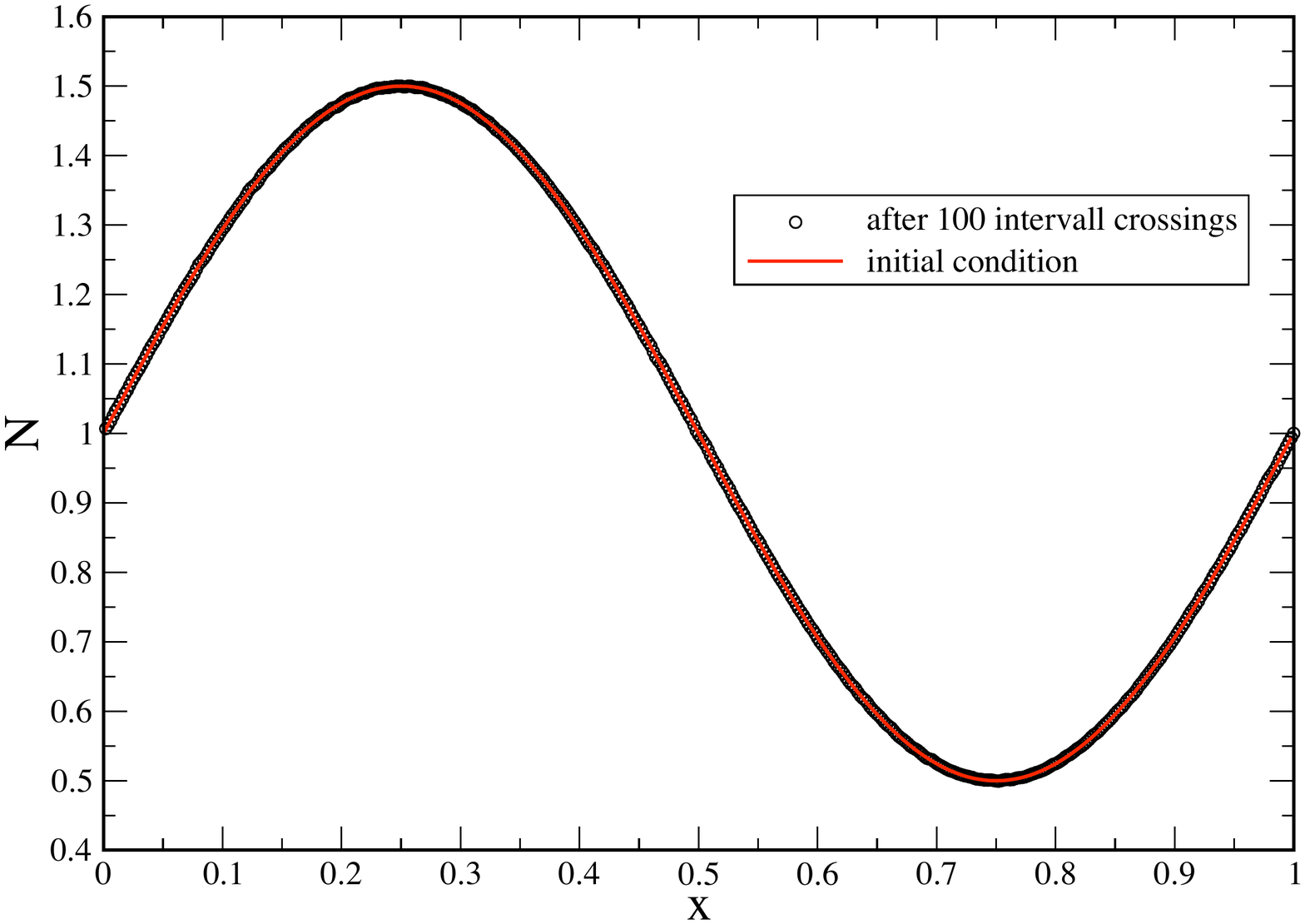} } 
   \caption{Advection of a relativistically moving sine wave (Lorentz factor $\gamma\approx 12.92$) across a periodic box: the initial condition of the computing frame number density, $N$, is shown as solid, red line, the result after as many as 100 box crossings is overplotted as circles.}
   \label{fig:sine_advection}
\end{figure}

In this test we explore the ability to accurately advect a smooth density pattern. 
We choose a sine wave that propagates towards the right through a 
periodic box. Since this test does not involve shocks, we switch off the artificial dissipation terms. 
We use 500 equidistantly placed SPH particles in the interval [0,1], enforce periodic 
boundary conditions and use a polytropic equation of state with $\Gamma=4/3$. We 
impose a computing frame number density $N(x)= N_0 + \frac{1}{2} \sin(2 \pi x)$, a 
constant velocity $v_0=0.997$ corresponding to a Lorentz factor $\gamma\approx 12.92$ and we instantiate a constant pressure corresponding to 
$P_0= (\Gamma-1) n_0 u_0$, where $n_0= N_0/\gamma$, $N_0=1$ and  $u_0=1$. The specific energies of the particles are chosen so that each particle has the same pressure $P_0$.\\
The advection of this relativistic sine wave is essentially perfect, see Fig.~\ref{fig:sine_advection}:
the result after as many as 100 box crossings (circles) is indistinguishable from the initial setup 
(red line), neither  wave amplitude nor phase have been noticeably affected during the evolution.

\subsubsection{Test 6: Relativistic advection of a square wave}
\begin{figure}[htbp] 
   \centerline{
   \includegraphics[width=5in]{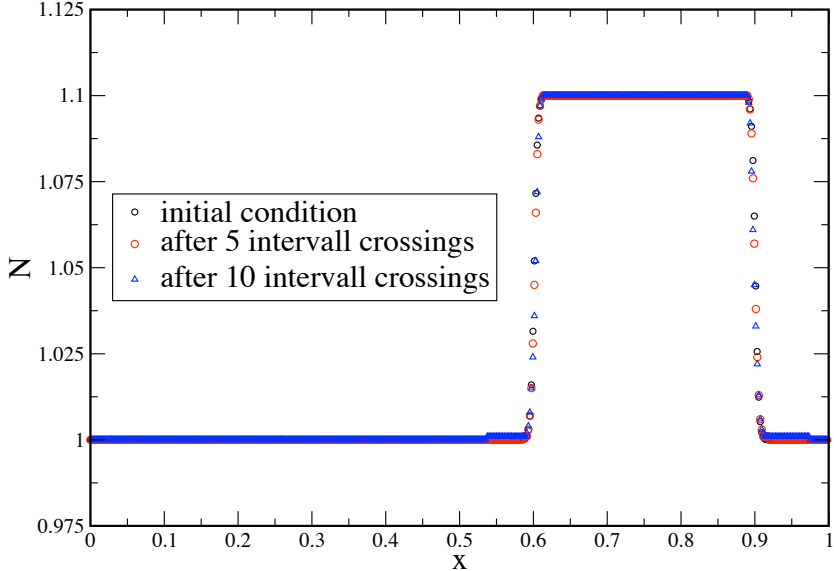}   } 
   \caption{Advection of a relativistically moving square wave (Lorentz factor $\gamma\approx 12.92$) across a periodic box: the initial condition of the computing frame number density, $N$, is shown as black circles, the result after 5  box crossing times (red circles) and 
   10 box crossing times (blue triangles) are overplotted. }
   \label{fig:square_advection}
\end{figure}

In this test we advect a square wave through the interval [0,1], again using periodic boundaries.
Due to the involved steep flanks this problem is substantially more challenging than the previous
one. We represent our box-shaped number density profile, $N(x)$, numerically as the sum of two 
Fermi-functions transiting from a lower state $N_{\rm low}$ at $x_1$ to a higher state, $N_{\rm high}$,
and back to the lower state at $x_2$:
\begin{equation}
N(x)= N_{\rm low} + (N_{\rm high}-N_{\rm low}) \left\{\frac{1}{1+\exp(\frac{x-x_2}{\Delta x})} - \frac{1}{1+\exp(\frac{x-x_1}{\Delta x})} \right\},
\end{equation}
where $\Delta x$ sets the length scale on which the transitions occur. For the numerical experiments, we use $N_{\rm low}= 1.0$, $N_{\rm high} =1.1$ and $\Delta x$ is set to twice the particle spacing in the low density region. We use equal mass particles in this test, constant pressure throughout the 
box is instantiated as in the sine wave problem above and we impose again a constant initial velocity of $v_0= 0.997$.\\
The results are displayed in Fig.~\ref{fig:square_advection}: the black circles show the initial number
density profile, the results after 5 and 10 box crossings are shown as red circles and blue triangles, respectively. Overall the results of this challenging test agree very well with the initial state, but after 10 box crossings the flanks have slightly softened, and the low density state close to the flanks
has been slightly increased. 

\subsubsection{Test 7: Evolution of a relativistic simple wave}
\begin{figure}[htbp] 
 \vspace*{-3.3cm}
            \centerline{
\hspace*{0.57cm}\includegraphics[width=4.2in]{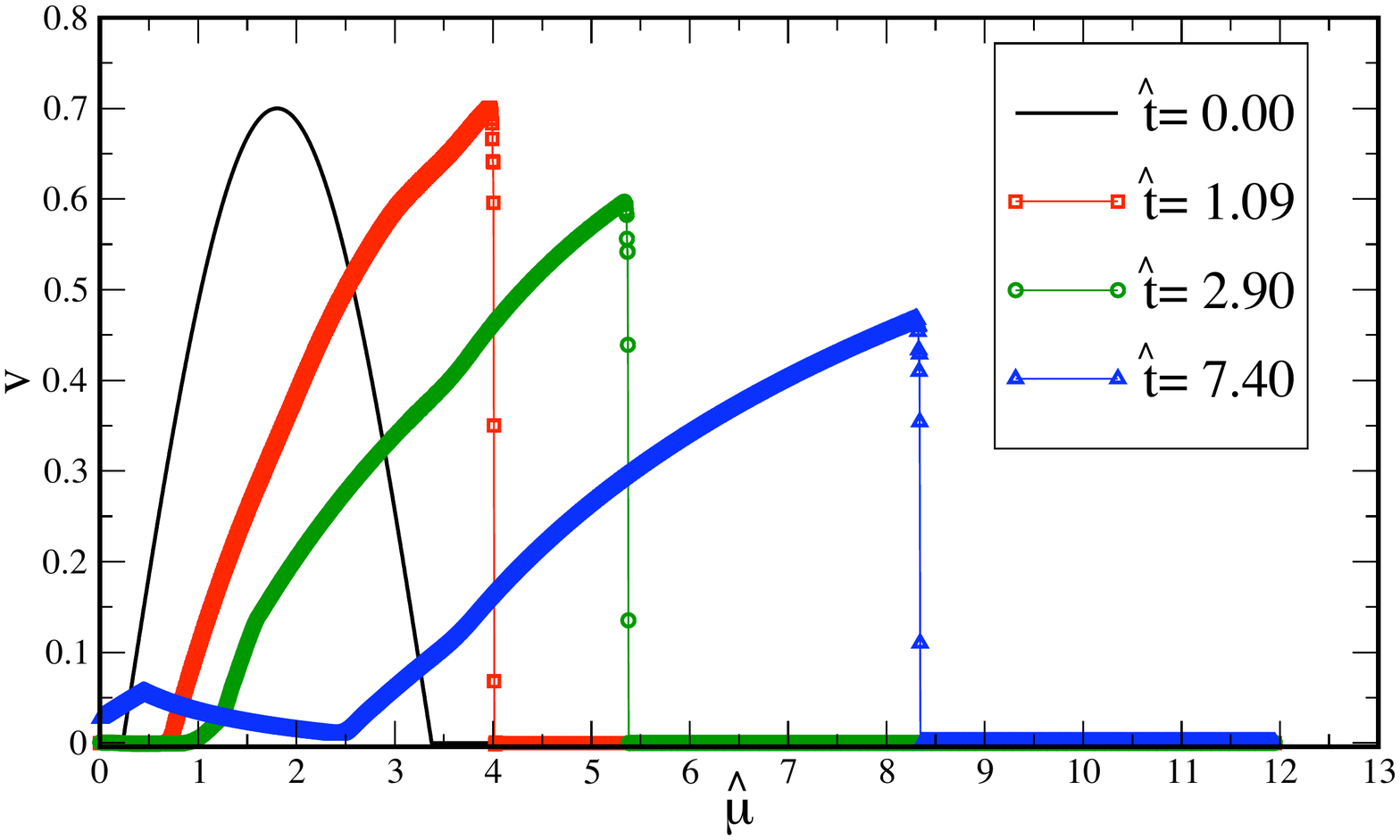} \hspace*{-1.4cm}     
   \includegraphics[width=4.2in]{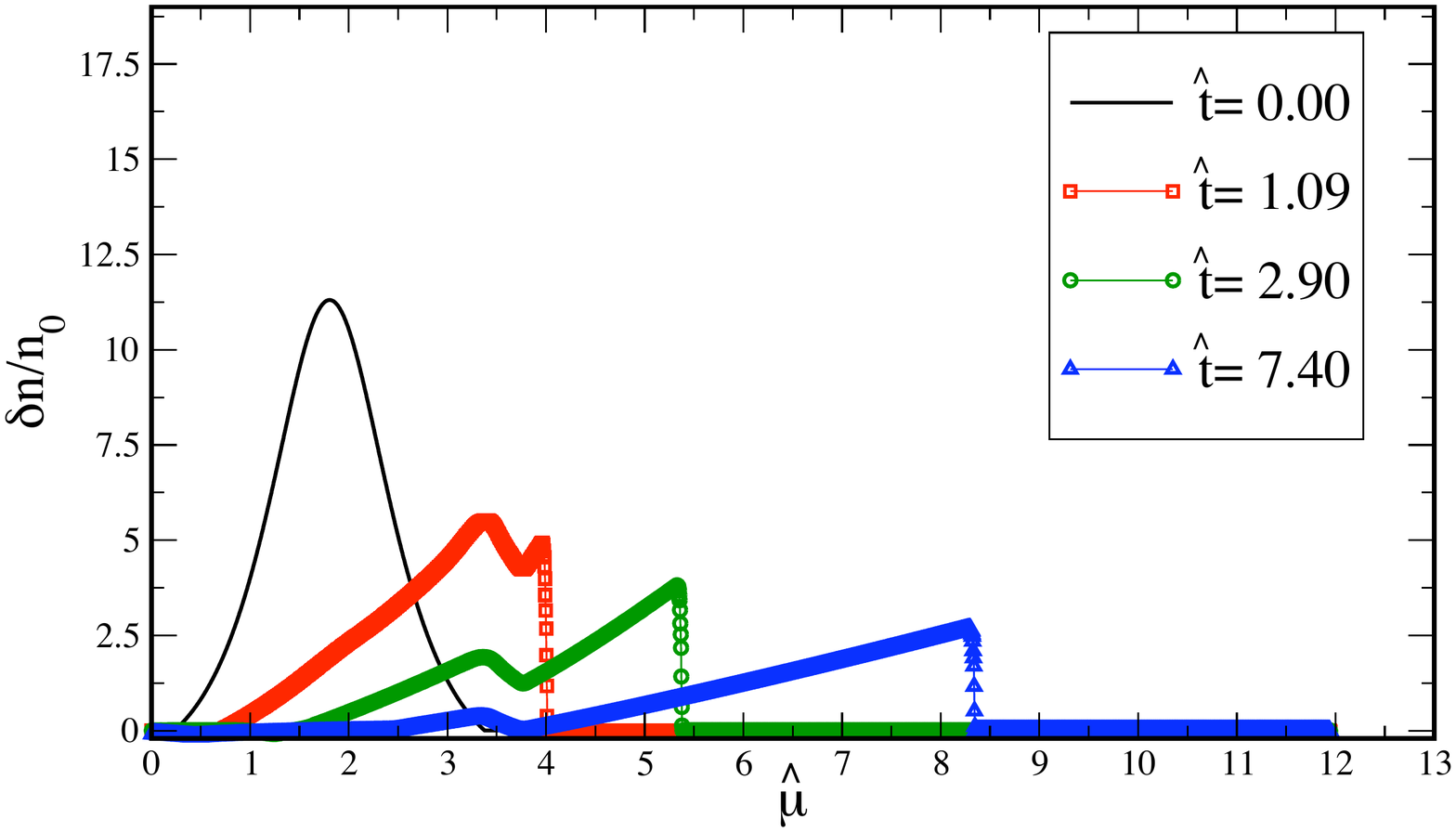}}  
   
    \vspace*{-1.6cm}
     \centerline{  \hspace*{0.7cm}   \includegraphics[width=4.2in]{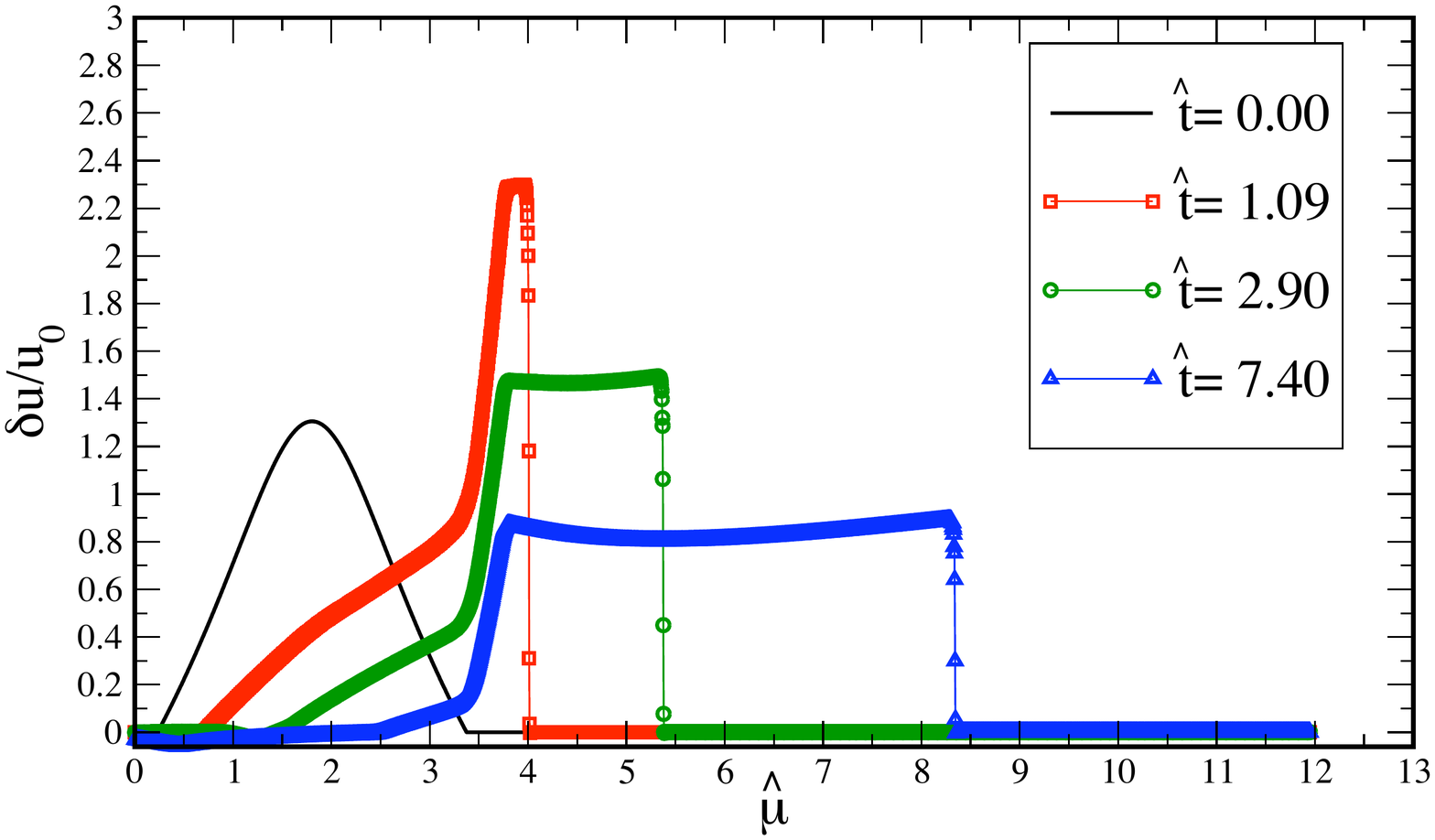}  \hspace*{-1.4cm}       
      \includegraphics[width=4.2in]{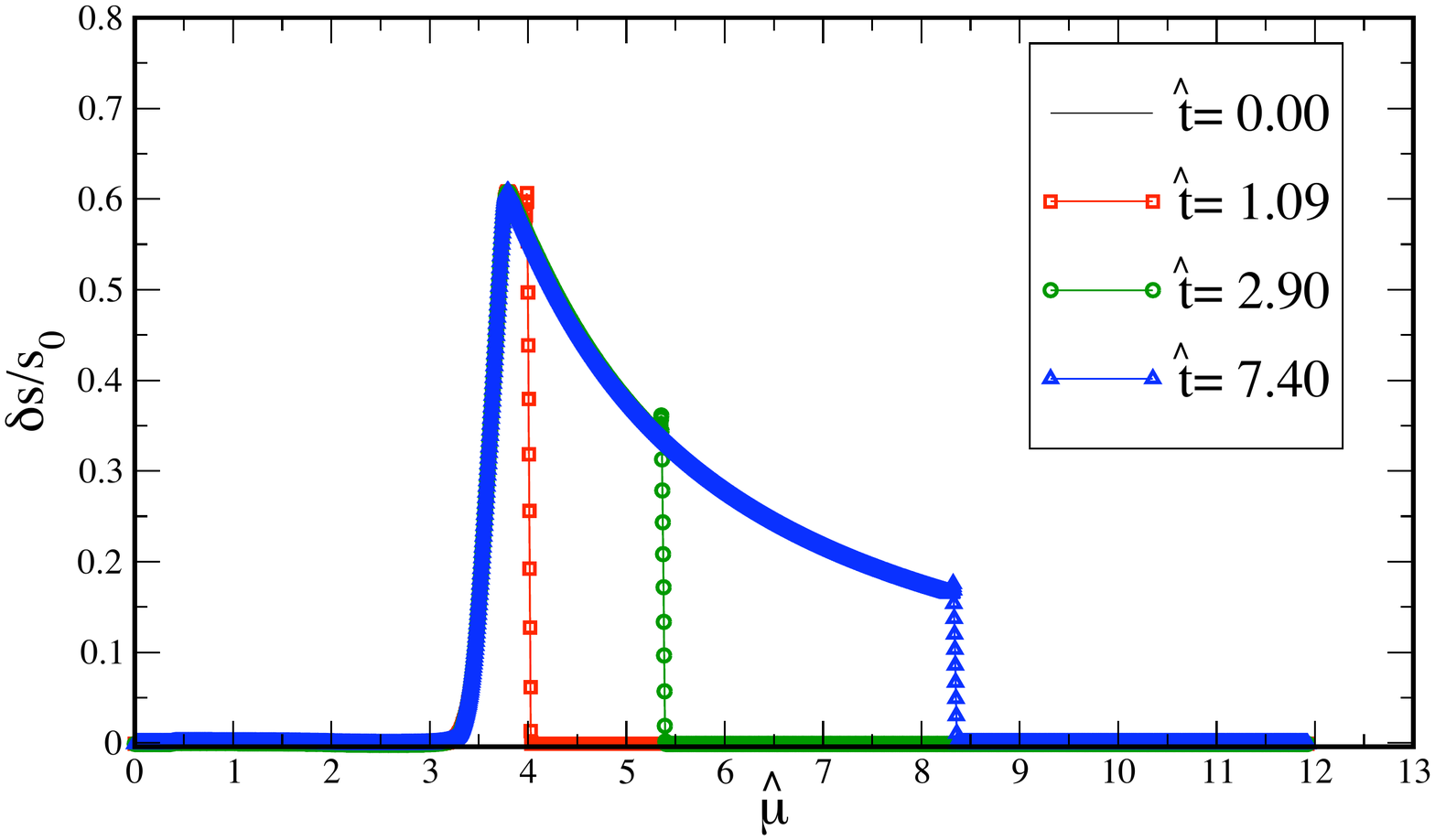} }    
      \vspace*{-1.5cm}
   \caption{Evolution of a relativistic simple wave: starting from a velocity sine pulse propagating to
   the right into a uniform medium, a shock forms that subsequently dissipates the wave. Shown are
   the velocity (upper left), and the contrasts in (rest frame) density (upper right), specific energy (lower left) and entropy (lower right).}
   \label{fig:simple_wave}
\end{figure}

Here we present results of a challenging test that involves relativistic simple waves
and is rarely shown in the literature.\\
Relativistic simple waves \cite{taub48,eltgroth71,liang77,anile83,anile89} are characterized by
the spatial and temporal constancy of two of the three Riemann invariants for one dimensional 
fluid flows. The three Riemann invariants are the specific entropy, $s$, and the quantities
\begin{equation}
J_{\pm}= \ln(\gamma + U) \pm \int \frac{c_{\rm s}}{\rho} d\rho,
\end{equation}
where $c_{\rm s}$ is the sound speed and $U$ the x-component of the four velocity. Simple waves propagating to the right/left are characterized by the constancy of $s$ and $J_{-}/J_{+}$.  In this
test, a purely compressive initial sine pulse is set up that propagates into a static, uniform medium. 
According to the results of \cite{anile83}, the initial velocity pulse will steepen while maintaining its 
peak velocity until it evolves into a relativistic strong shock. From thereon, 
the wave will dissipate and continuously decrease in velocity and in the contrasts of density and internal
energy (as measured with respect to the initial unperturbed state).\\
Our setup and parameter choice closely follows  \cite{anile83}.  We use the equation of state of a 
radiation-dominated fluid, $p= k(s) \rho^{4/3}$. The initial 
state consists of an unperturbed fluid state denoted by subscript $0$ with an overlaid  "velocity pulse".
Like in \cite{anile83}, we choose the sound velocity in the unperturbed state as $c_0=0.3$ and display normalized, dimensionless quantities. For the initial (local rest frame) density we choose $n_0= 1$
and we use the length of our initial velocity pulse, $l_0=100$, as characteristic length scale. The normalized quantities are denoted with a $\hat{  }$ -symbol: 
$\hat{\mu}= \mu/(n_0 l_0)$, where $\mu$ is the mass coordinate, $\hat{n}= n/n_0$, $\hat{u}= u/u_0$,
$\hat{s}= s/s_0$, $\hat{P}=P/P_0$ and $\hat{t}=t/l_0$.
We proceed in the following steps: 
\begin{itemize}
\item choose a sinusoidal velocity profile with a maximum $v_{\rm max}= 0.7$ as a function of
$\hat{\mu}$. The width of the pulse in normalized mass coordinates is 
$\hat{\mu}_0= \pi$.  To keep the pulse sufficiently far away from the (fixed) boundaries particles (at $\hat{\mu}=0$ and 13), we choose $\hat{\mu}_{\rm peak}= 3$ as mass coordinate of the velocity maximum. For the ease 
of comparison, we plot the results with a constant $\hat{\mu}$-offset so that our initial setup (black curves in Fig.~\ref{fig:simple_wave}), coincides with the initial conditions of \cite{anile83}, see their Figure
5.
\item Once $v$ is known, we calculate the sound velocity \cite{liang77,anile83} as a function of $\hat{\mu}$
\begin{equation}
c_{\rm s}= \frac{\frac{1+c_0\sqrt{3}}{1-c_0 \sqrt{3}} \left( \frac{1+v}{1-v}\right)^{\frac{1}{2 \sqrt{3}}} - 1}     {\sqrt{3} \left[ \frac{1+c_0\sqrt{3}}{1-c_0 \sqrt{3}} \left( \frac{1+v}{1-v}\right)^{\frac{1}{2 \sqrt{3}}} + 1\right]}
\end{equation}
\item and from this the specific energy and rest frame number density
\begin{equation}
u= \frac{9 c_{\rm s}^2}{4(1-3c_{\rm s}^2)} \quad {\rm and} \quad n= n_0 \left(\frac{u}{u_0}\right)^3.
\end{equation}
\item Finally, the computing frame baryon number density, $N= \gamma n$, is calculated which, in turn, allows to assign positions and baryon numbers.
\end{itemize}

We display in Fig.~\ref{fig:simple_wave} $v$, $\delta n/n_0= \hat{n}-1$, $\delta u/u_0= \hat{u}-1$ and $\delta s/s_0= \hat{s}-1$, where the axis limits and output times are chosen as in  Fig. 5 of \cite{anile83}. The specific entropy is nowhere used in our SPH formulation, but can be post-processed  for the chosen equation of state as $\hat{s}= \hat{P}^{3/4} \hat{n}^{-1}$.

Generally, we find very good agreement with the results of \cite{anile83}. The initial sine-pulse continuously steepens while keeping its maximum velocity constant (to within about 2\%) until
a relativistic shock forms (at $\hat{t}\approx 1$), see Fig.~\ref{fig:simple_wave} upper left. Note in particular that the positions of the shock fronts agree excellently with those found in \cite{anile83}.
The wave subsequently dissipates, thereby producing a double peak structure in the density contrast
(at $\hat{t}= 1.09$, upper right) with the dip coinciding with the peak in the entropy (lower right) and the 
steep flank at $\hat{\mu}= 3.5$ in the specific energy (lower left). The entropy shows a small, spurious overshoot at the leading edge which is the result dividing $\hat{P}^{3/4}$ and $\hat{n}$ at the shock front, but apart from this, the agreement with \cite{anile83} is very good.\\
We only find minor differences. Our simulations are less dissipative, the second density peak, for example, still exceeds the leading one at $\hat{t}= 1.09$ and is still visible at 7.40 while by the same time
it has vanished in \cite{anile83}. Moreover, our specific energy  at $\hat{t}= 1.09$ shows a clear plateau
behind the shock while theirs is a very smooth peak (which may just be the result of lower resolution),
and our entropy peak (at $\hat{\mu} \approx 3.75$) is slightly smaller than theirs ($\delta s/s_0= 0.61$ vs. $\approx 0.67$).

\subsection{Tests in 2D}
Multi-dimensional calculations are complicated by the fact that the SPH particles have the
possibility to pass each other, which can cause imperfect particle distributions and numerical noise
(mainly in the velocity). And in fact, noise is a major concern for multi-dimensional SPH calculations. 
Our strategy in this respect is threefold: a) since they can be a major reason for noise, we take particular 
care to prepare accurate initial conditions, b) we have implemented an artificial dissipation 
scheme  that --in addition to shocks-- also triggers on velocity noise, see Sect. \ref{sec:AV_controle}, and iii)
we use relatively large values for the parameter $\eta$ in Eq.~(\ref{eq:dens_summ_SR_N_b}).\\
\subsubsection{Initial particle distribution}
\label{sec:2D_initial_particle_dist}
\begin{figure}[htbp] 
   \centerline{\includegraphics[width=8cm]{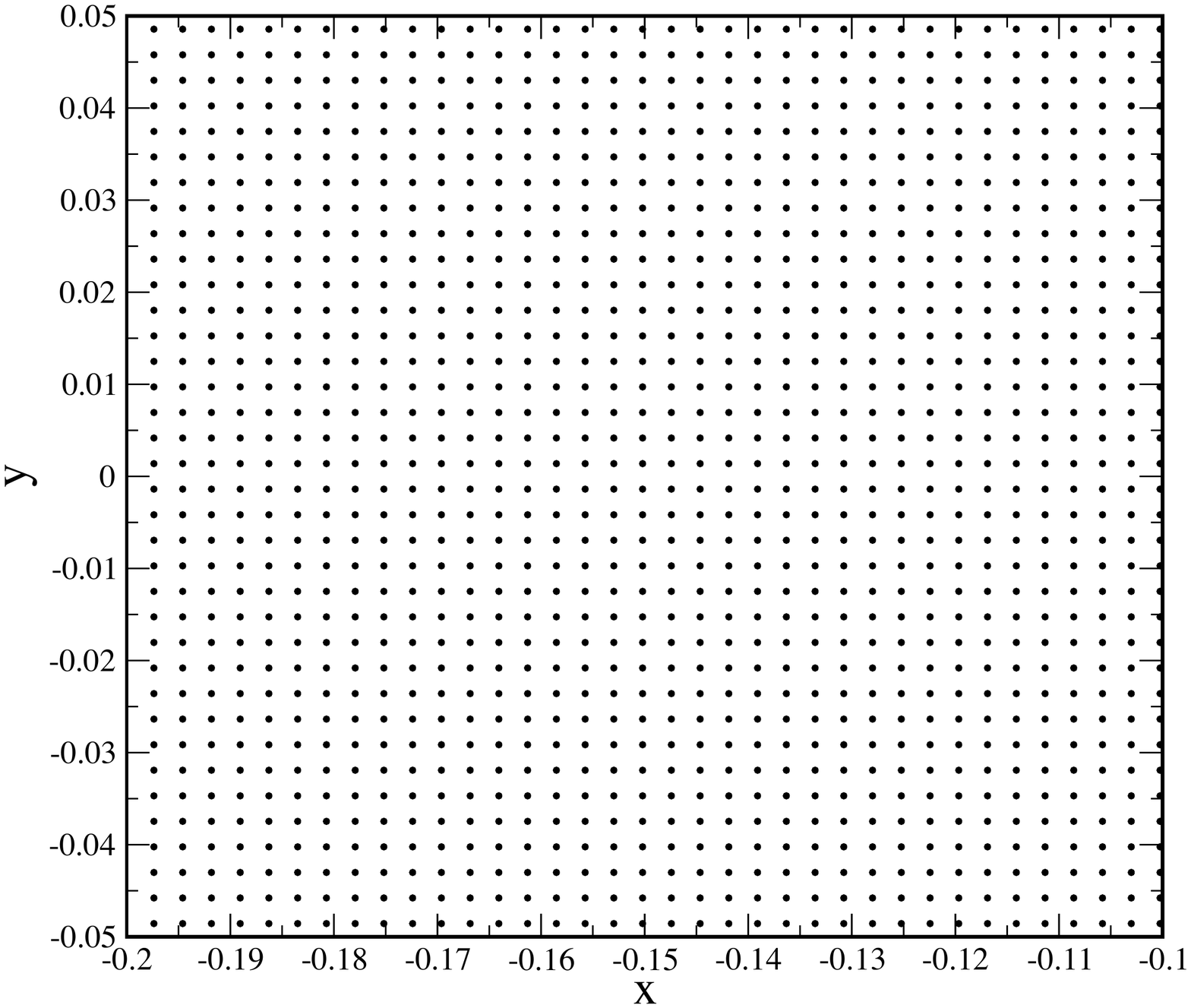} \hspace*{-1cm}
                       \includegraphics[width=8cm]{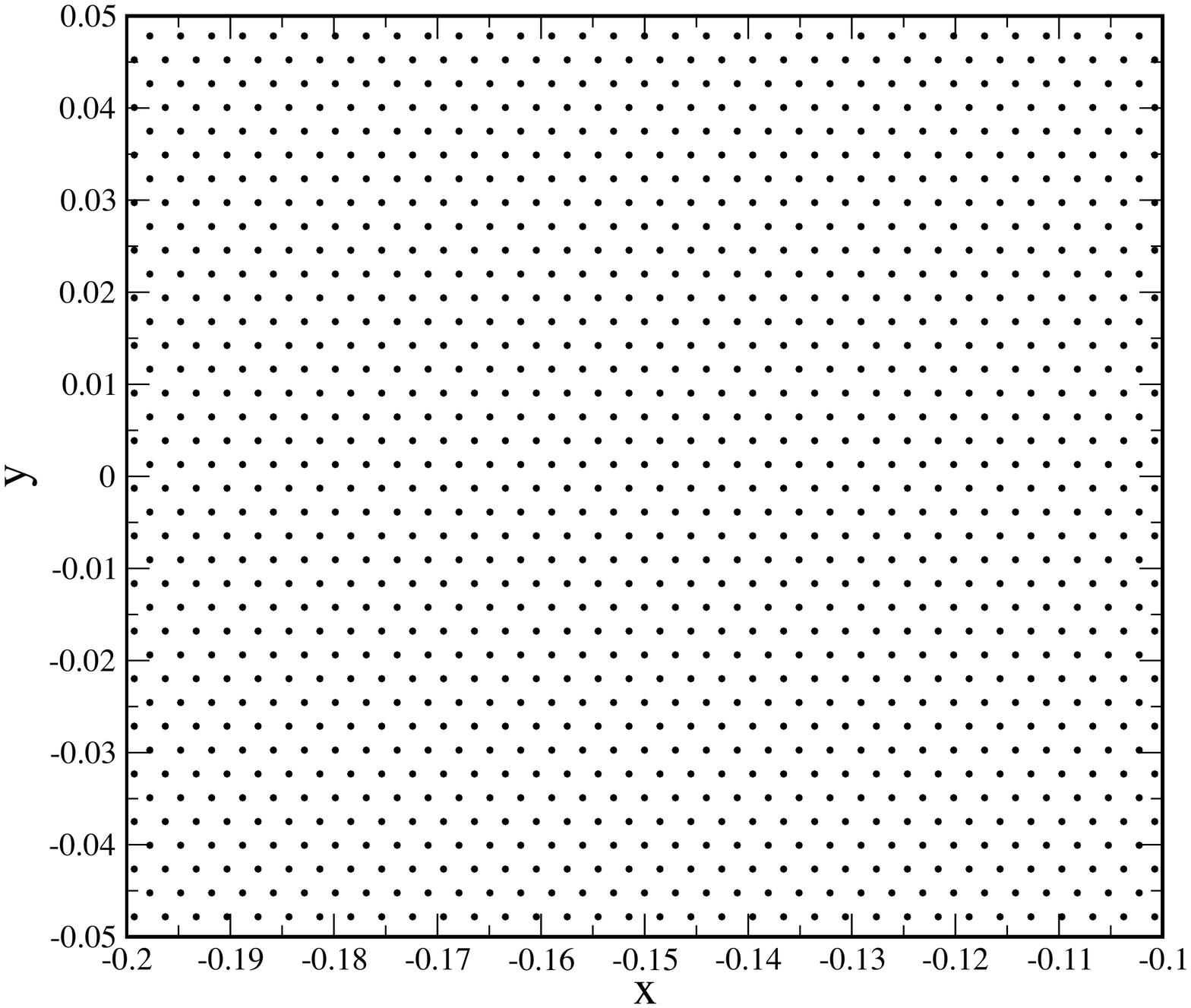}}
   \caption{Initial particle distributions: grid (left) and hexagonal lattice (right).}
   \label{fig:grid_CP_distrib}
\end{figure}

\begin{figure}[htbp] 
   \centering
   \centerline{\hspace*{0.4cm}\includegraphics[width=6.3cm]{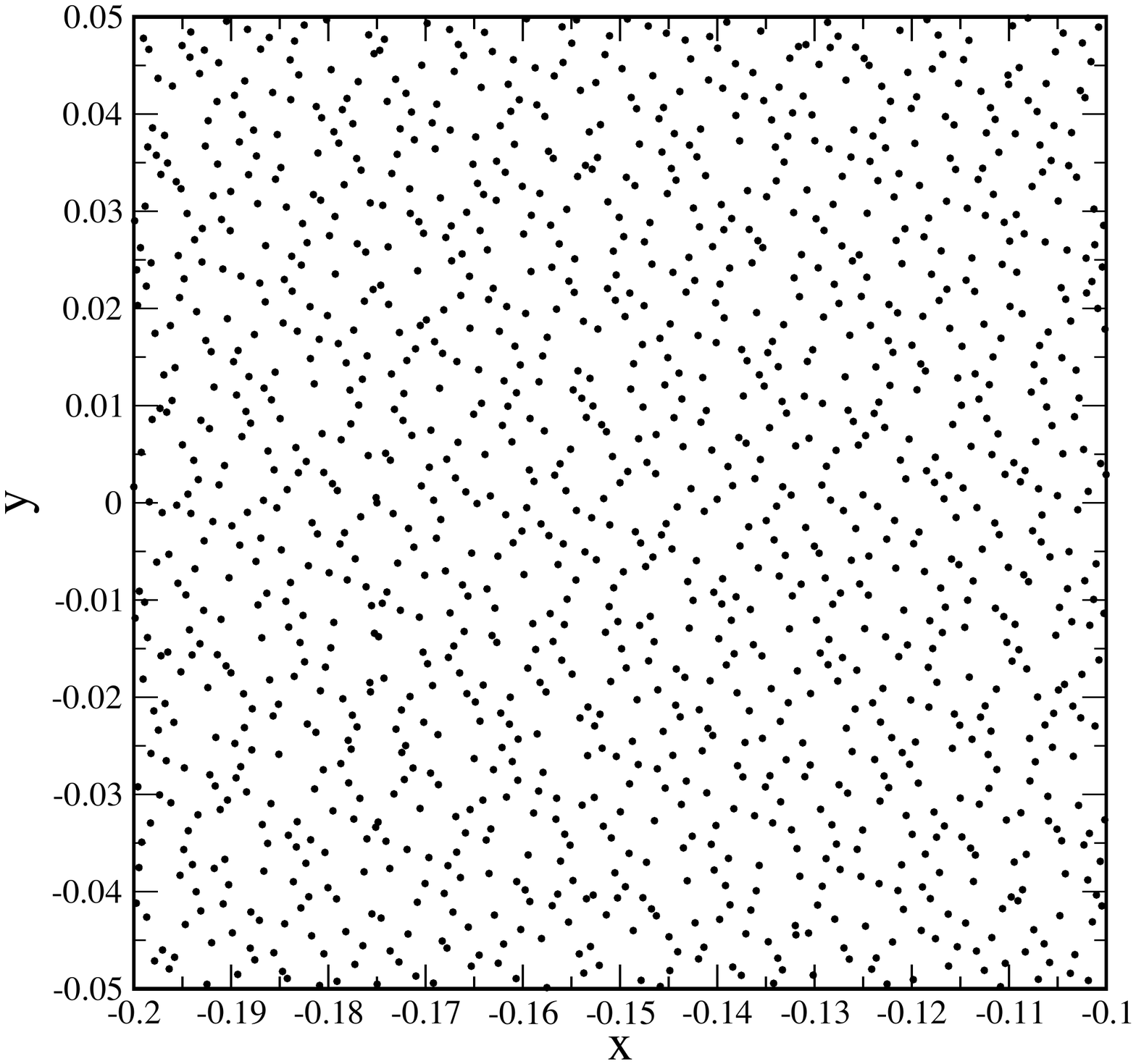} \hspace*{-1.5cm}\includegraphics[width=6.3cm]{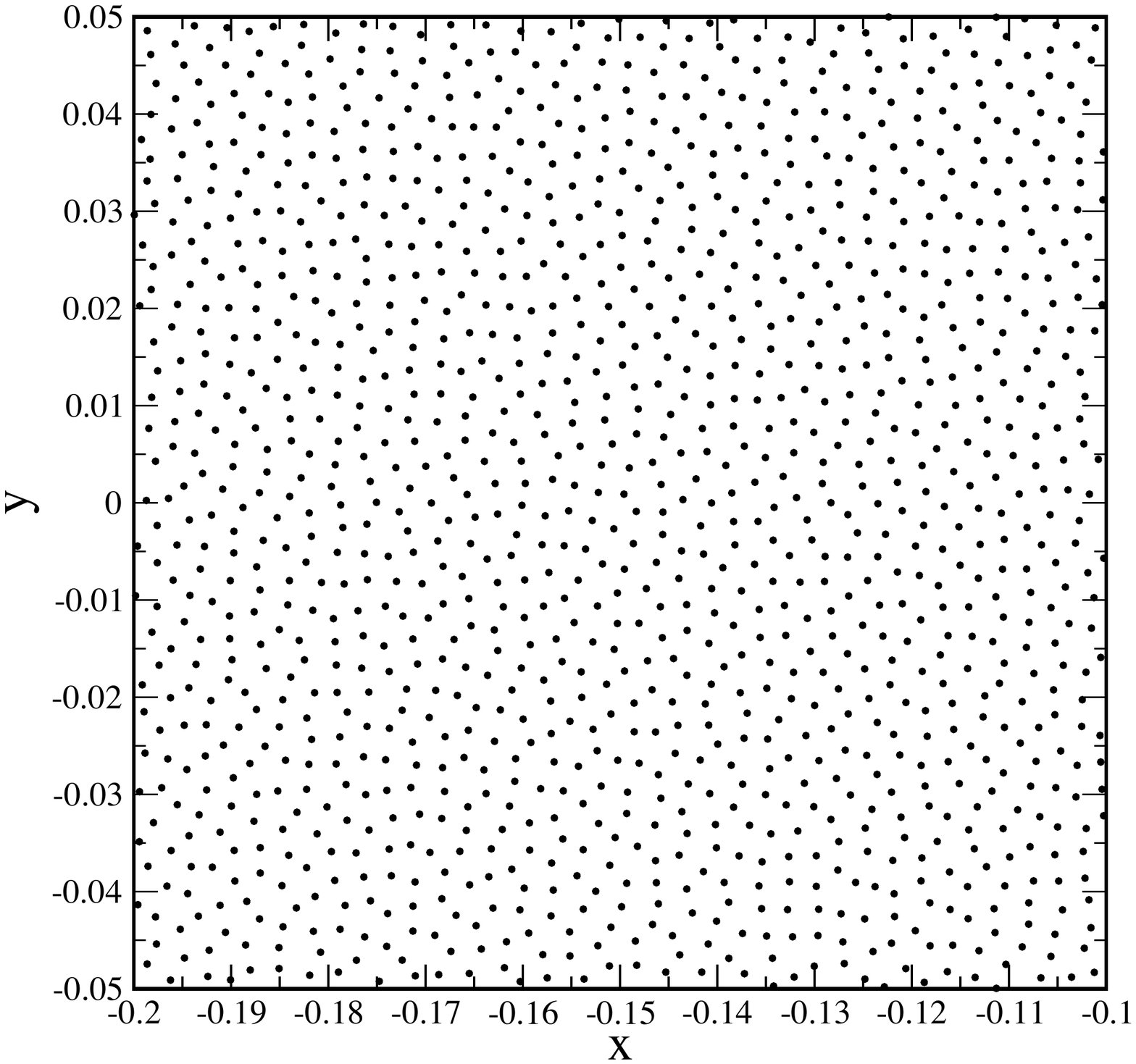}   \hspace*{-1.5cm}\includegraphics[width=6.3cm]{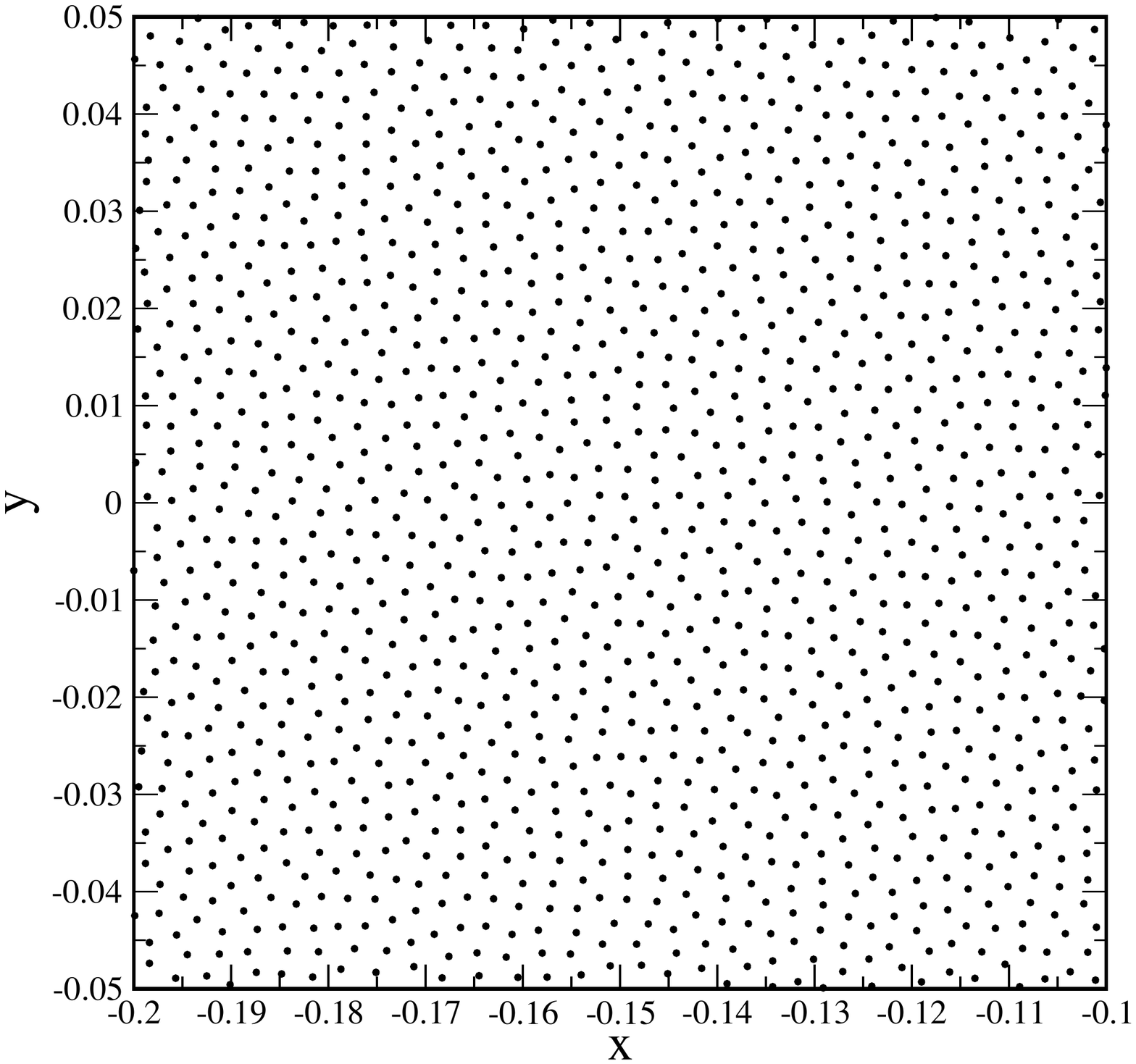}}
       \caption{Particle distributions: initial, quasi-random Sobol sequence (left), after regularization sweeps (middle) and
       final configuration after relaxation (right). In this process the peaked LIQ kernel was used.}
   \label{fig:particle_distributions}
\end{figure}

We have performed some experiments with the initial particle setup. We explored three types of initial 
particle distributions: i) a perfect equidistant grid, Fig.~\ref{fig:grid_CP_distrib}, left, ii) a hexagonal 
lattice, corresponding to the distribution of the centers of close-packed spheres, Fig.~\ref{fig:grid_CP_distrib}, 
right, and iii) a "glass"-like particle distribution obtained in a relaxation process.  
To produce the "glass-like" distribution,  we proceed in three steps: initially, the particles are distributed 
according to a Sobol quasi-random sequence \cite{press92}, see Fig.~\ref{fig:particle_distributions}, left panel. 
In a second step, we choose an "adjustment  time step", $\Delta t_{\rm{adj},a}= 0.5 \;  (h_a/c_{{\rm s},a})$, 
and perform several sweeps where we use the force law
\be
\vec{f}_a= - \sum_b \frac{\nu_b}{N_b} P_b \nabla W_{ab}(h_a),
\ee 
which is a very simple discretization of the Euler equation. In each loop, the particle positions are updated according to
\be
\vec{r}_a= \vec{r}_a + \frac{1}{2} \; \vec{f}_a \; \Delta t_{\rm{adj},a}^2,
\ee
while the maximum value of the force $|\vec{f}_a|$ is constantly monitored. This is done for as long as the 
maximum force value is decreasing from one sweep to the next. Typically,  the maximum force is reduced by 
two orders of magnitude with respect to the initial Sobol sequence and provides a visually more regular 
particle distribution, see middle panel in Fig.~\ref{fig:particle_distributions}. Since the particles do 
not move much during the optimization process,  we store for each particle a (generous) candidate list 
which is used for all the sweeps. Therefore, this procedure is a computationally  inexpensive way to drive 
the particles into a regular distribution. To further optimize the distribution, we "relax" this particle 
distribution to an optimal state by applying a large dissipation constant, $K=3$, to the momentum, but not 
the energy equation (we want to obtain a perfect particle distribution, but not artificially heat the 
system). During this process, particles at the edges and, for shock tubes, near the transition between 
the two states, are kept fix, so that each state relaxes separately.\\
In our tests we have found the best results with particles initially placed on the hexagonal lattice. The
(somewhat artificial) particle distribution on a equidistant grid gave good shock tube results at low resolution, 
but at higher resolution produced a lot of velocity noise once the particles 
leave the grid. This effect was more pronounced for the peaked LIQ kernel, which we still prefer for the 2D tests 
since it performed slightly better in the below tests than the standard CS kernel.

\subsubsection{Test 8: 2D relativistic shock tube}
\label{sec:2D_shocks}
\begin{figure}[htbp]
\centering 
   \includegraphics[width=16cm]{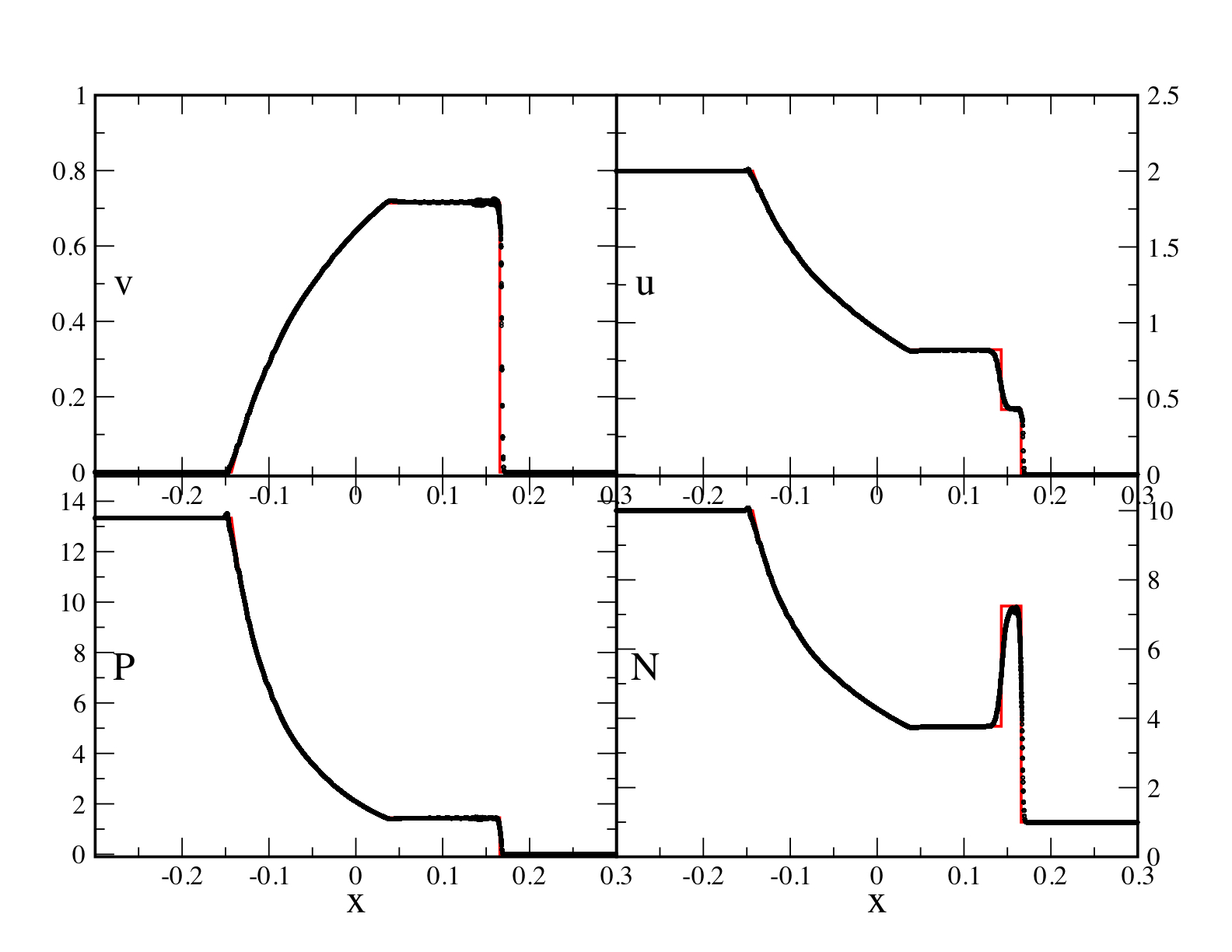} 
     \caption{2D relativistic shock tube test (140 000 particles). Values at individual particle positions (i.e. no averaging or smoothing has been applied) are shown as black circles, the exact solution is indicated by the red line.}
   \label{fig:ST_I_2D}
\end{figure}

\begin{figure}[htbp] 
   \centering
   \centerline{\includegraphics[width=16cm]{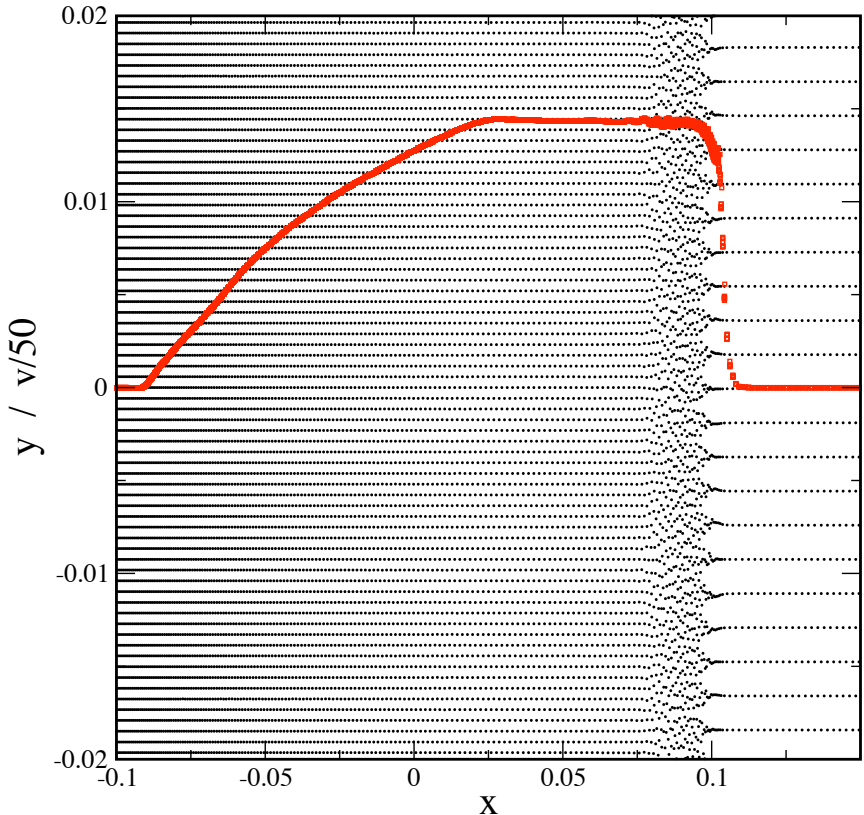} }
       \caption{Zoom into the particle distribution around the shock front in a 2D relativistic shock tube test.
                       Shown are the particle positions (black dots) and the scaled particle velocities, $v_a/50$, as red squares. Initially the particles were distributed on a hexagonal lattice.}
   \label{fig:zoom_shock}
\end{figure}

\begin{figure}[htbp] 
   \centering
   \centerline{\includegraphics[width=14cm]{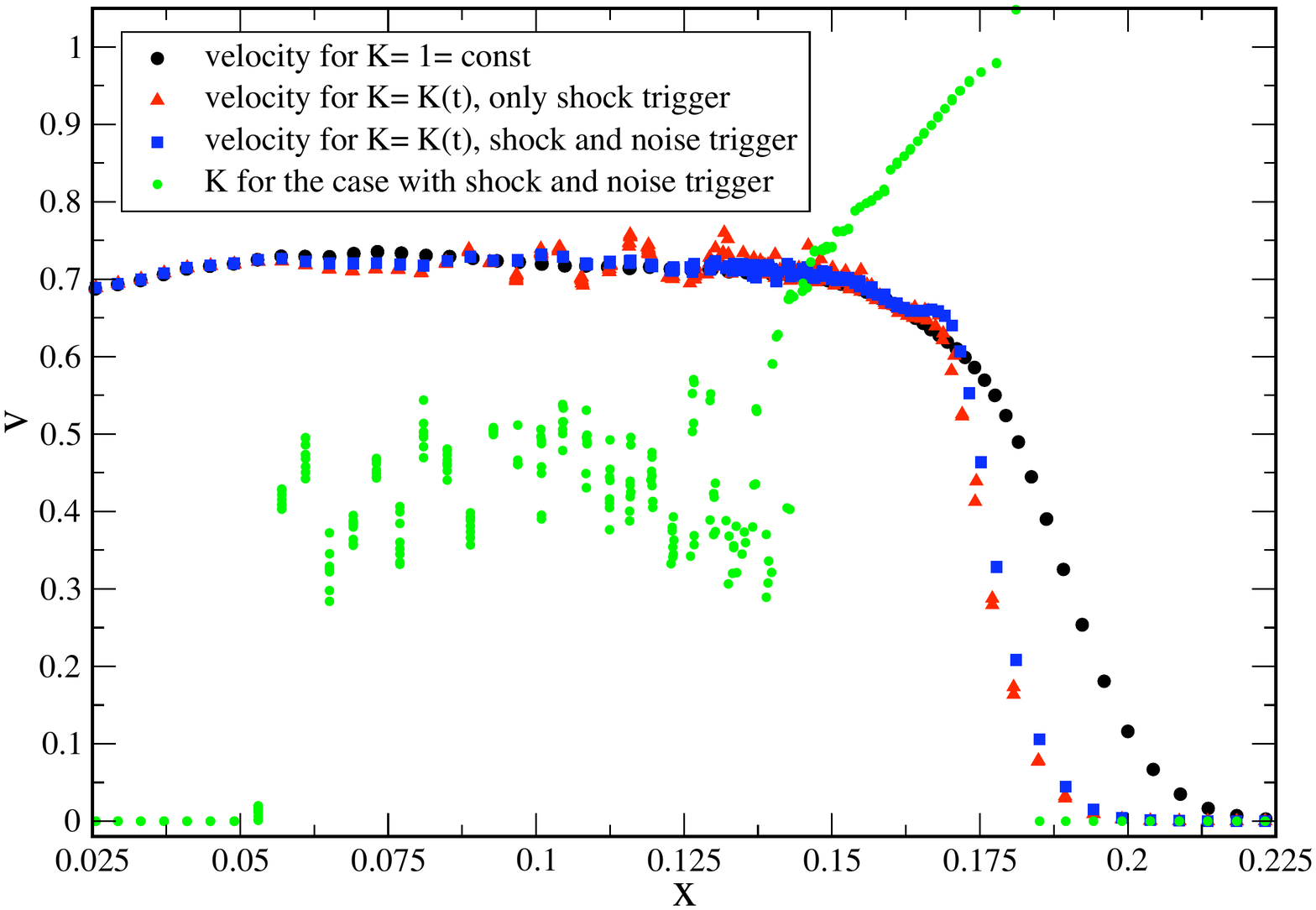} }
       \caption{Zoom into the shock region of a 2D, low-resolution (2000 particles), relativistic shock 
        tube test. The velocities resulting from the new dissipation control scheme are shown as blue 
        squares, the used values for the dissipation parameter $K$ as green circles. The red triangles
         denote the results from the same scheme, but with the noise trigger switched off. The results 
         obtained with constant dissipation parameter $K=1$ are shown as black circles.}
   \label{fig:comparison_full_shock_only_constant}
\end{figure}

\begin{figure}[htbp] 
   \centering
   \centerline{\hspace*{0.4cm}\includegraphics[width=9.5cm]{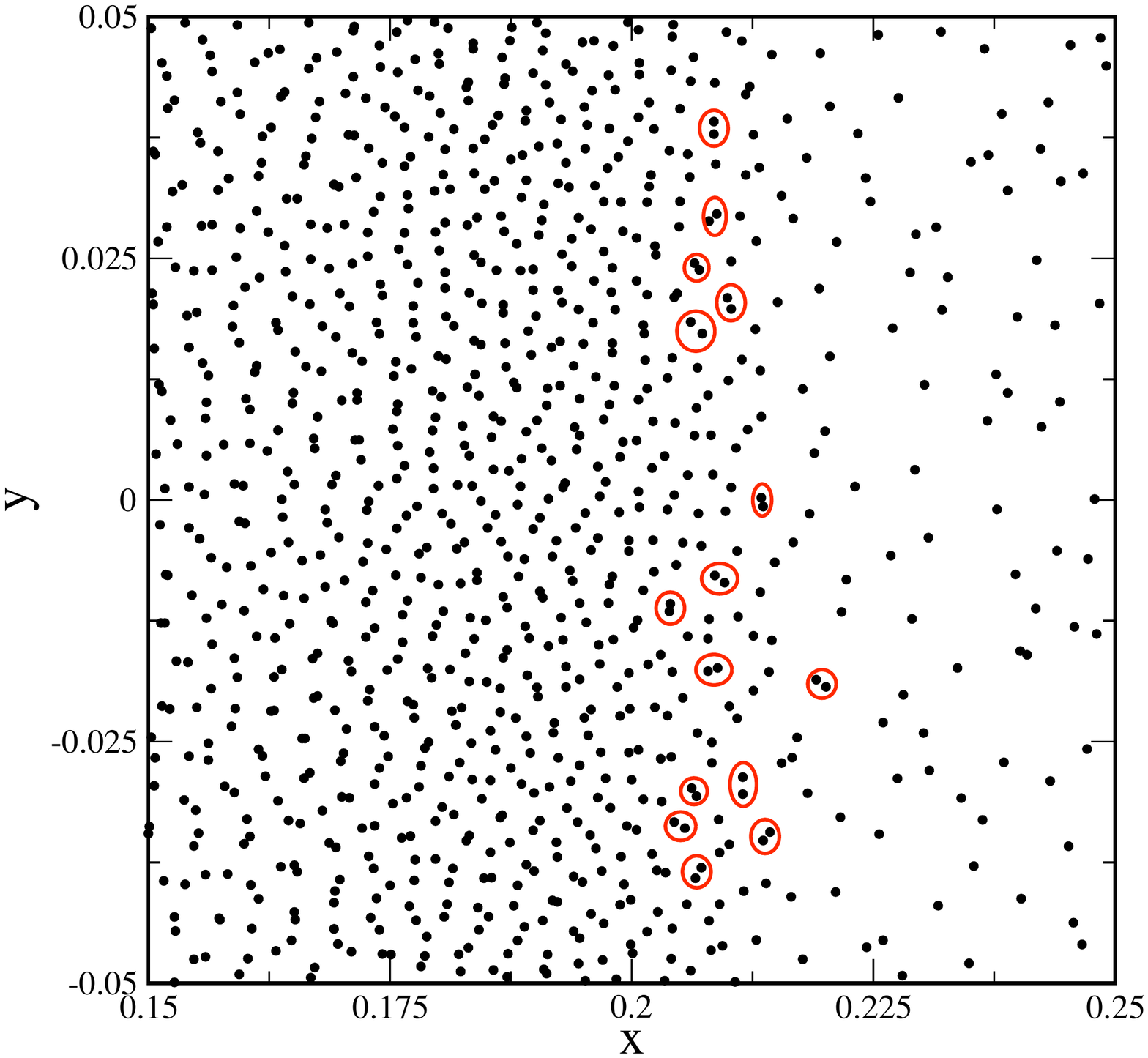} \hspace*{-1.9cm}\includegraphics[width=9.5cm]{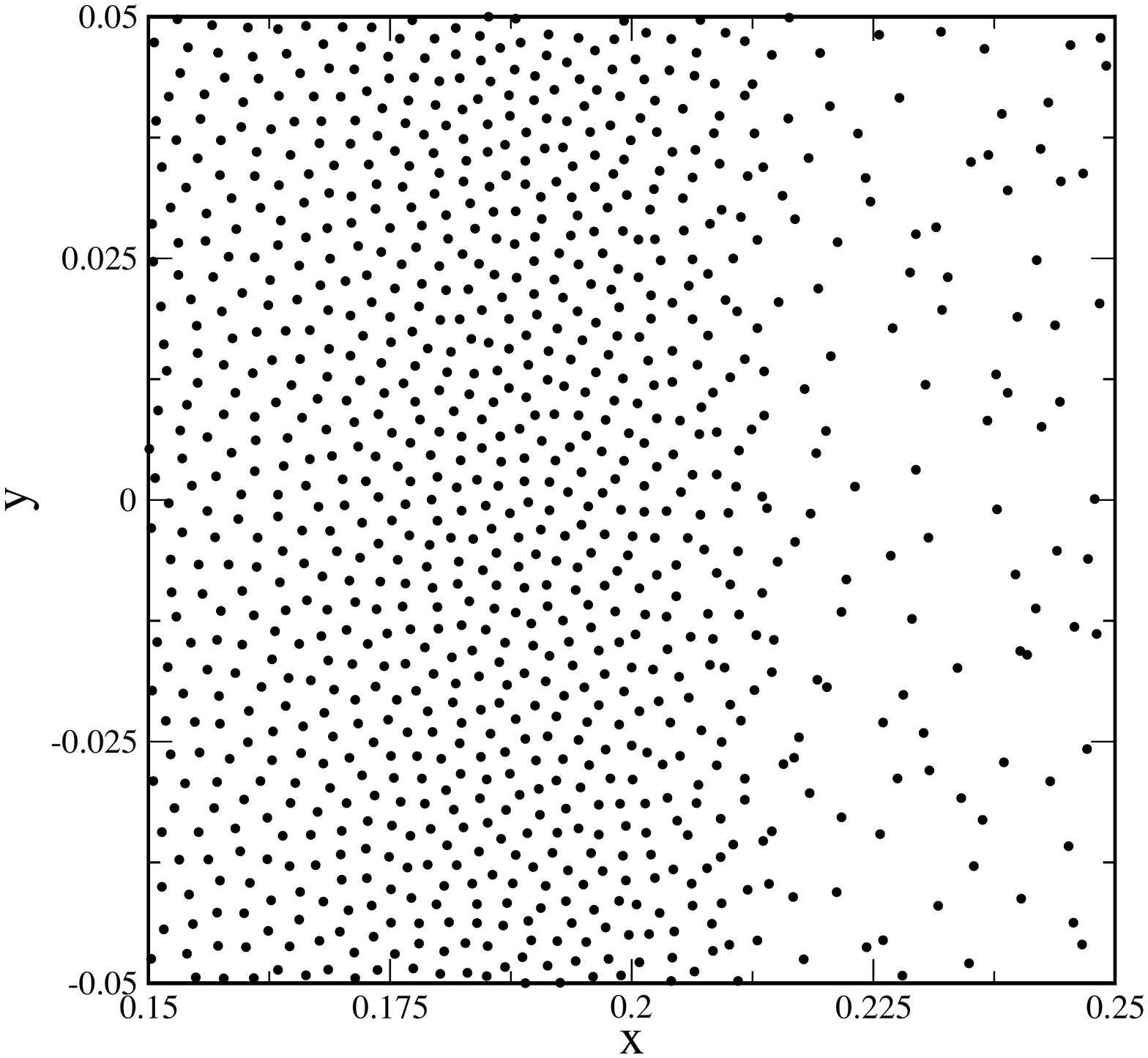}}
       \caption{Low-resolution ("glass") comparison of the cubic spline (CS; left) and the Linear 
                    Quartic (LIQ; right) kernel for a 2D, relativistic shock tube test. The left panel shows a 
                    substantial fraction of "paired" SPH particles (some of which are highlighted with red 
                    ellipses). The peaked LIQ kernel (right panel), in contrast, only produces a few pairs 
                    directly at the shock front, but the post-shock region is very regularly sampled again.}
   \label{fig:pairing}
\end{figure}

\begin{figure}[htbp] 
   \centering
   \centerline{\includegraphics[width=8cm]{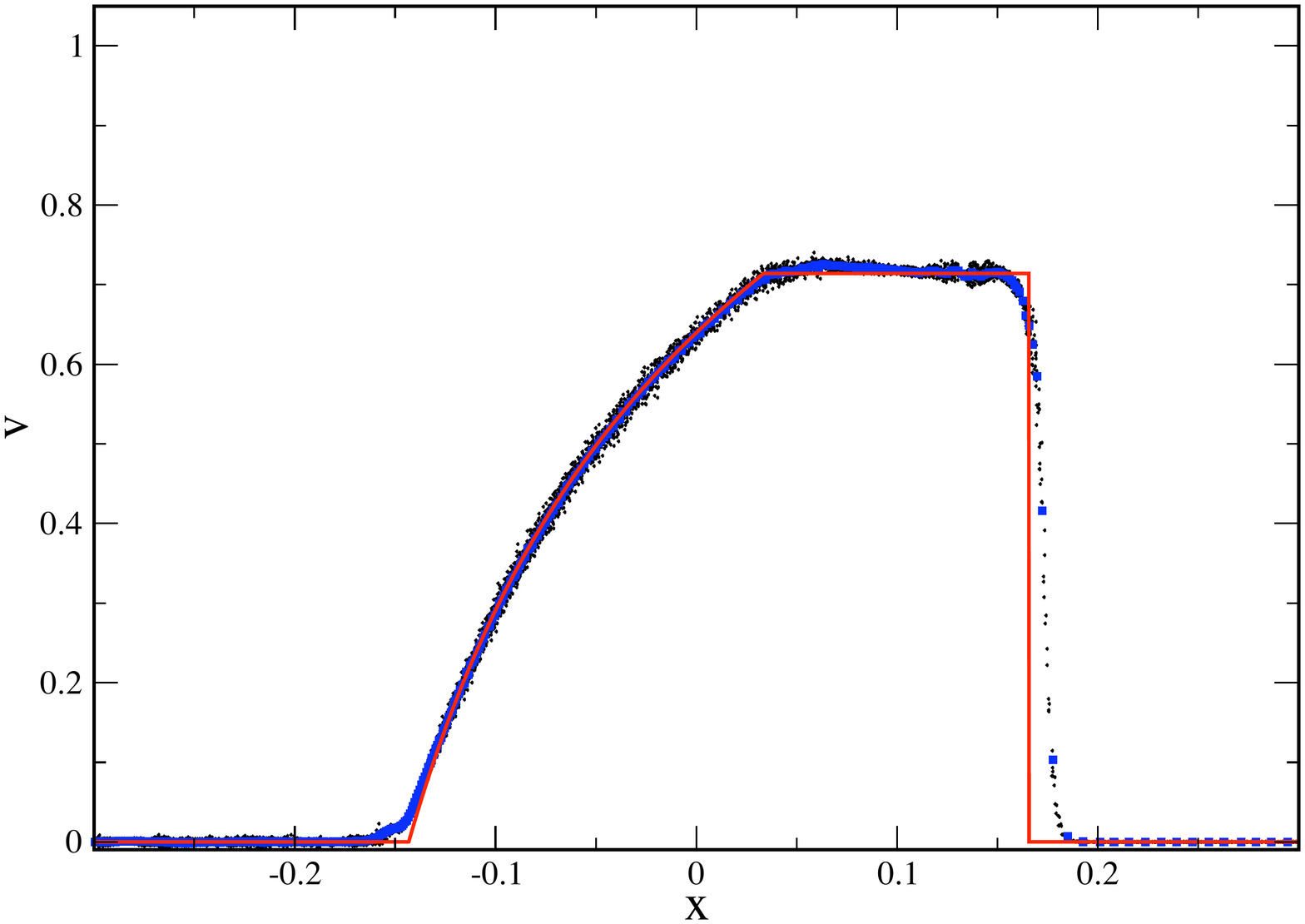} \hspace*{-1cm} \includegraphics[width=8cm]{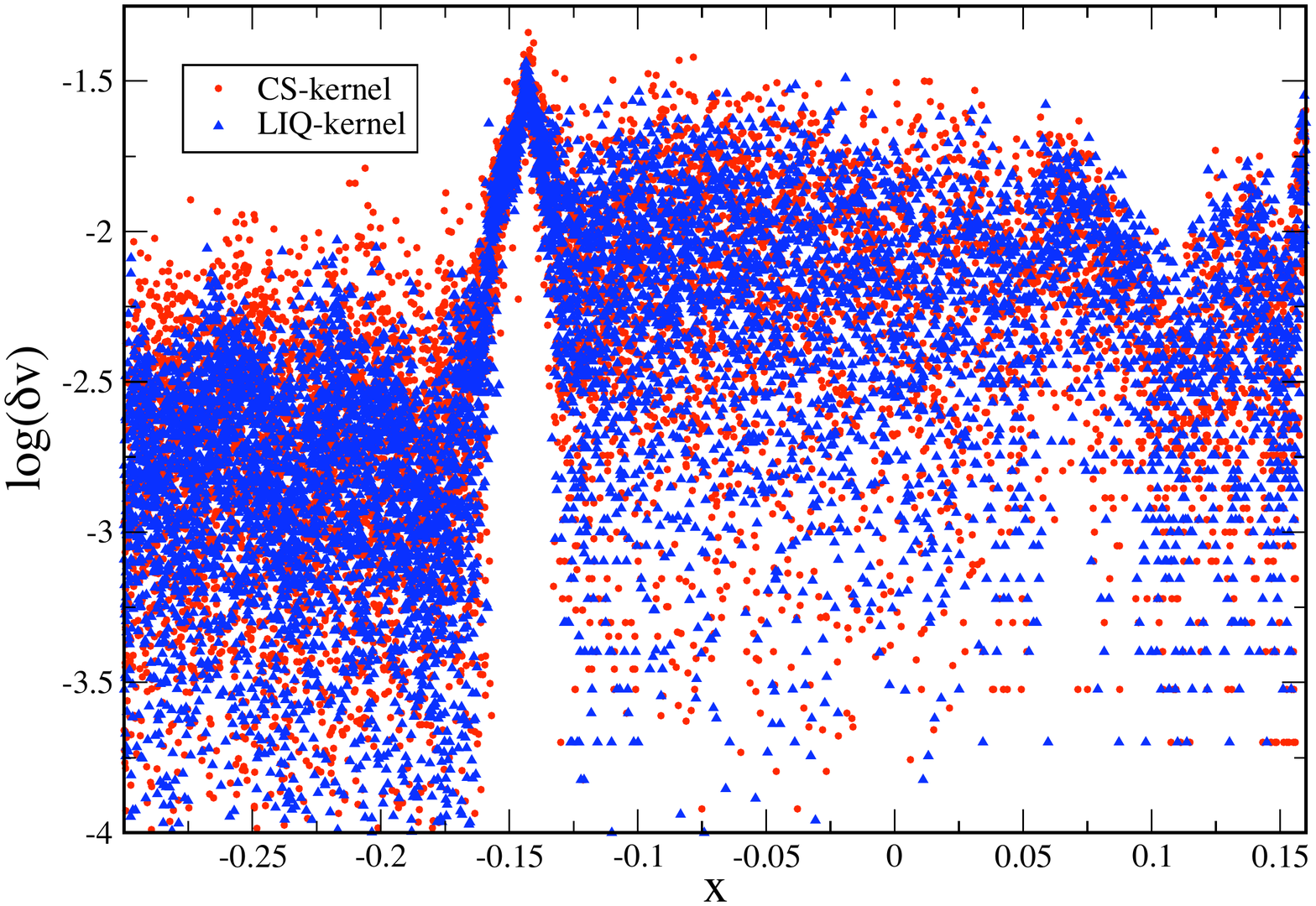} }
             \caption{Low-resolution, 2D shock simulation starting from an imperfect initial particle distribution.
                          Left: velocity for the  linear quartic kernel (black circles), average velocity (blue square) and exact solution (solid red). Right: comparison of the velocity errors for the same test, once with the CS
                          (red circles) and once with the LIQ kernel (blue triangles).}
   \label{fig:shock_CS_vs_LIQ}
\end{figure}

This is the 2D version of test 1 shown above, i.e. the initial conditions are $(N,v,P)^{\rm L}=(10,0,40/3)$ 
and $(N,v,P)^{\rm R}=(1,0,10^{-6})$. For this test we place 140 000 particles on a hexagonal lattice 
between [-0.4,0.4] $\times$ [-0.02,0.02] and use our standard 2D parameter set: 
$\eta= 1.7, K_{\rm max}= 2., K_{\rm min}= 0., A_{\rm ref, shock}= 0.8, A_{\rm ref, noise}= 5$ together with the LIQ
kernel. The results are displayed in Fig.~\ref{fig:ST_I_2D} with SPH particle properties as black circles 
and the exact solution as red line. Generally, we find a 
very good agreement with the exact solution, only the contact discontinuity is somewhat smeared out, similar 
to the 1D case. Some small high-frequency oscillations in the velocity occur behind the shock front. They are 
mainly caused by the particles that under the action of the peaked kernel have to change from the low density 
lattice into a higher density lattice. This is illustrated in Fig.~\ref{fig:zoom_shock} for a lower resolution 
calculation (40 000 particles) where we have plotted the re-scaled velocities over the particle distribution. 
The small post-shock irregularities in the velocity are clearly related to the transition region behind the 
shock front at $x\approx 0.08$. They could be further reduced at the expense of applying more dissipation, but 
we consider the current parameter set as a good compromise between low dissipation and absence of noticeable 
post-shock oscillations.\\
In the subsequent low-resolution tests (2000 particles), we illustrate the performance of the dissipation control scheme.
We perform all tests with $\eta= 1.7, K_{\rm max}= 2., K_{\rm min}= 0., A_{\rm ref, shock}= 0.8, A_{\rm ref, noise}= 5.$ and the 
LIQ kernel. The results from the full scheme are shown in Fig.~\ref{fig:comparison_full_shock_only_constant} as  blue 
squares. Note that the parameter $K$ (green circles) remains at zero up to the arrival of the shock front where it 
jumps to values close to unity. In the post-shock region it remains around 0.5 (triggered by noise), and vanishes again in the expansion 
fan. For comparison, we perform another test with identical parameters but with $K= 1=$ const everywhere (black circles), 
and one more where we switch off the noise trigger (red triangles). Clearly, the new scheme substantially sharpens 
shock fronts, essentially without compromising the post-shock region. With shock trigger only, the shock is sharp, but 
substantial post-shock oscillations occur. The two independent triggers allow to apply a different amount of dissipation
to both phenomena.\\
We also perform two tests to gauge the influence of the smoothing kernel, in the first one, we want to explore to 
which extent "pairing" of SPH particles occurs, in the second, we start from an imperfect initial particle 
distribution and explore the influence of the kernel on the resulting velocity noise.  The result of the 
first experiment is shown in Fig.~\ref{fig:pairing} where we zoom into the shock fronts of a low resolution 
test, once with the CS and once with the LIQ kernel. In both cases the initial particle distribution is 
obtained by the above described relaxation process. The CS kernel (left) produces many particle pairs 
(some are highlighted by red ellipses) which deteriorates the volume sampling of the SPH particles. The 
LIQ kernel (right), in contrast, produces only temporarily very few pairs directly at the shock front, but 
the post-shock region is again sampled very regularly.\\
In the second test, we start from an imperfect initial particle distribution (10 000 SPH particles in 2D) 
as produced by the regularization sweeps, see above. The particles are subsequently assigned the properties 
that correspond to the left and right state and finally evolved, once with the CS and once with the LIQ kernel. 
The results at $t=0.2$ are shown in Fig.~\ref{fig:shock_CS_vs_LIQ}. The imperfect initial conditions introduce 
some scatter in the velocities (left), but the average values still agree very well with
the exact solution. To assess the performance of the different kernels in this situation, we plot in 
Fig.~\ref{fig:shock_CS_vs_LIQ}, right panel, the deviation of the particle velocities, $\vec{v}_a$, from the 
exact solution, 
$\vec{v}_{\rm ex}$,
\be
\delta v_a \equiv |\vec{v}_a - \vec{v}_{\rm ex}(\vec{r}_a)|.
\ee
The LIQ kernel (blue triangles) produces noticeably smaller errors than the "standard" SPH kernel (red circles).\\
Summarizing our experiments with the two kernels, we find that the LIQ kernel performs slightly better,
it produces a substantially more regular particle distribution and yields smaller errors for initially noisy
particle configurations.

\section{Summary}

We have derived a new set of special-relativistic SPH equations from a variational principle. This
work differs from \cite{monaghan01} in accounting also for the special-relativistic ``grad-h'' terms, corrections
for usually neglected derivatives of the smoothing kernels with respect to the resolution lengths.
We have used an artificial viscosity prescription that is inspired by  Riemann solvers 
\cite{monaghan97}. We trigger independently on shocks and on velocity noise. Since the dissipation
applied in shocks is tuned according to the carefully measured density slope, shock fronts become 
substantially sharper than for a constant dissipation parameter, see Fig.~\ref{fig:comparison_full_shock_only_constant}. 
Contrary to earlier approaches \cite{morris97,rosswog00} we do not evolve
the dissipation parameter continously to the desired value, but instead increase it instantaneously, similar to the
approach of \cite{cullen10}. If not further triggered, the parameter subsequently decays to zero. None
of the triggers is specific to special relativity, both could be applied as well in Newtonian SPH.\\
We have carefully tested this new approach in a slew of numerical benchmark tests. We find that the 
relativistic grad-h terms increase the accuracy of the method, but usually only have a moderate effect.
The improvements are generally dominated by the new signal velocity and the time-dependent
viscosity parameters. The new approach yields excellent results in the numerical experiments. As expected 
for a purely Lagrangian scheme, it performs close to perfect in pure advection problems, even at large 
Lorentz factors, see test problems 5 and 6. What is more, it also yields accurate results even in very strong 
shock tests, which are usually considered a particular challenge for SPH. For example, the scheme is able to 
accurately handle wall shock problems with a Lorentz factor as large as $\gamma= 50 000$, see test 4. We 
also perform a rarely shown, very challenging test in which a relativistic simple wave steepens into a strong 
shock and subsequently dissipates (test 7). Our numerical results for this test are in close agreement 
with those of the original paper \cite{anile83}.
In a last set of tests, we have explored the performance in 2D, relativistic shocks. Again we find very good 
agreement with the exact solutions. In these multi-D tests we have also experimented with the peaked 
linear quartic kernel \cite{valcke10} which yields slightly better test results than the most commonly 
used cubic spline kernel. Numerical experiments \cite{rosswog10d} show that the scheme is 
second-order accurate for smooth flows and first-order accurate if shocks are involved.

{\bf Acknowledgements}
I want to thank SISSA (Trieste, Italy) and the Polytechnical University of Barcelona, UPC, for their
hospitality. It is a pleasure to acknowledge insightful discussions with Walter Dehnen, John Miller 
and Justin Read. \\
This work has been supported by DFG under grant RO 3399/5-1.












\bibliography{astro_SKR}
\bibliographystyle{elsart-num-sort.bst}






\end{document}